\documentclass[aps,prl,twocolumn,superscriptaddress,shopacs]{revtex4}
\usepackage{graphicx}
\usepackage{bm}
\usepackage{epstopdf}
\usepackage{multirow}
\usepackage{float}
\usepackage{tabularx}
\usepackage{color}
\usepackage{xr}
\usepackage{amsmath}
\usepackage{color}

\begin{document}

\title{Short-range nematic fluctuations in Sr$_{1-x}$Na$_x$Fe$_2$As$_2$ superconductors}

\author{Shan Wu}
\email{shanwu@berkeley.edu}
\affiliation{Department of Physics, University of California, Berkeley, California, 94720, USA}
\affiliation{Material Sciences Division, Lawrence Berkeley National Lab, Berkeley, California, 94720, USA}

\author{Yu Song}
\email{yusong@berkeley.edu}
\affiliation{Department of Physics, University of California, Berkeley, California, 94720, USA}
\affiliation{Material Sciences Division, Lawrence Berkeley National Lab, Berkeley, California, 94720, USA}

\author{Yu He}
\affiliation{Department of Physics, University of California, Berkeley, California, 94720, USA}

\author{Alex Frano}
\affiliation{Department of Physics, University of California, San Diego, California 92093, USA}

\author{Ming Yi}
\affiliation{Department of Physics and Astronomy, Rice University, Houston, Texas 77005, USA}

\author{Xiang Chen}
\affiliation{Department of Physics, University of California, Berkeley, California, 94720, USA}
\affiliation{Material Sciences Division, Lawrence Berkeley National Lab, Berkeley, California, 94720, USA}

\author{Hiroshi Uchiyama}
\affiliation{Japan Synchrotron Radiation Research Institute, SPring-8, 1-1-1 Kouto, Sayo, Hyogo 679-5198, Japan}

\author{Ahmet Alatas}
\affiliation{Advanced Photon Source, Argonne National Laboratory, Argonne, Illinois 60439, USA}

\author{Ayman H. Said}
\affiliation{Advanced Photon Source, Argonne National Laboratory, Argonne, Illinois 60439, USA}

\author{Liran Wang}
\affiliation{Institute for Quantum Materials and Technologies, Karlsruhe Institute of Technology, 76021 Karlsruhe, Germany}

\author{Thomas Wolf}
\affiliation{Institute for Quantum Materials and Technologies, Karlsruhe Institute of Technology, 76021 Karlsruhe, Germany}

\author{Christoph Meingast}
\affiliation{Institute for Quantum Materials and Technologies, Karlsruhe Institute of Technology, 76021 Karlsruhe, Germany}

\author{Robert J. Birgeneau}
\email{robertjb@berkeley.edu}
\affiliation{Department of Physics, University of California, Berkeley, California, 94720, USA}
\affiliation{Material Sciences Division, Lawrence Berkeley National Lab, Berkeley, California, 94720, USA}

\date{\today}
\begin{abstract}
Interactions between nematic fluctuations, magnetic order and superconductivity are central to the physics of iron-based superconductors. Here we report on in-plane transverse acoustic phonons in hole-doped Sr$_{1-x}$Na$_x$Fe$_2$As$_2$ measured via inelastic X-ray scattering, and extract both the nematic susceptibility and the nematic correlation length. By a self-contained method of analysis, for the underdoped ($x=0.36$) sample, which harbors a magnetically-ordered tetragonal phase, we find it hosts a short nematic correlation length $\xi\sim10$~$\rm \AA$ and a large nematic susceptibility $\chi_{\rm nem}$. The optimal-doped ($x=0.55$) sample exhibits weaker phonon softening effects, indicative of both reduced $\xi$ and $\chi_{\rm nem}$. 
Our results suggest short-range nematic fluctuations may favor superconductivity, placing emphasis on the nematic correlation length for understanding the iron-based superconductors. 

\end{abstract}

\maketitle
Unconventional superconductivity in the iron-based superconductors (FeSCs) appears near a putative quantum critical point where collective excitations could play an important role in the pairing mechanism.  These fluctuations and corresponding orders emerge from the interplay between lattice, spin and orbital degrees of freedom. Nematic order, in which rotational symmetry is broken while preserving translational symmetry, is a prominent example of such behavior \cite{Fernandes_2012}.  Abundant evidence indicates that the corresponding nematic fluctuations are closely connected to unconventional superconductivity in the FeSCs \cite{Kuo2016,fradkin2010,fernandes2014,Chu2012}.

\begin{figure*}
\includegraphics[width=2\columnwidth,clip,angle =0]{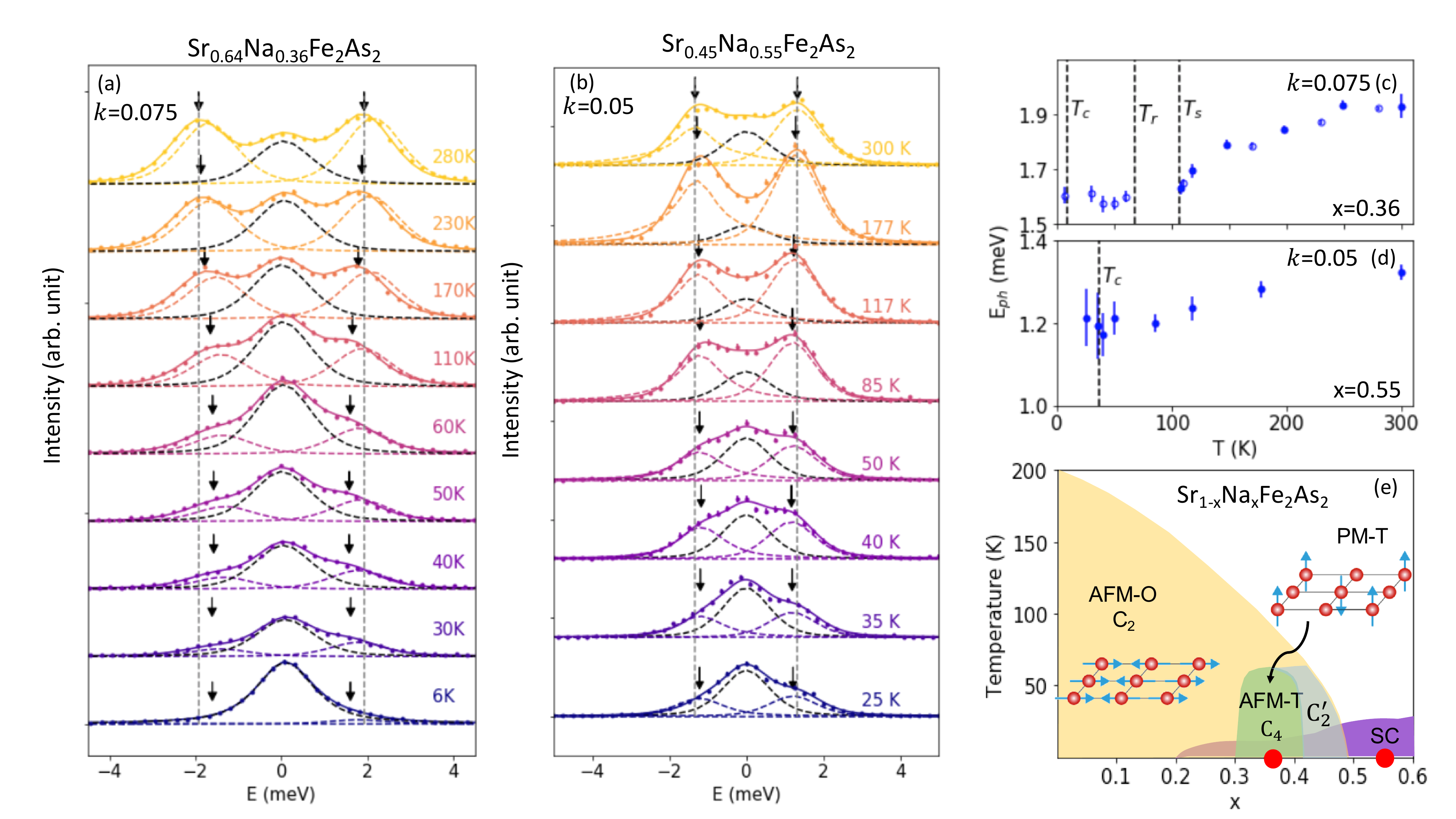}
\caption{Representative temperature-dependent energy scans for (a) Sr$_{0.64}$Na$_{0.36}$Fe$_2$As$_2$ at wave vector ${\bf Q}=(2,k,0)$ with $k$ = 0.075 from SPring-8 and (b)  Sr$_{0.45}$Na$_{0.55}$Fe$_2$As$_2$ with $k  = 0.05$ from APS. Solid symbols and dashed curves are experimental data and the corresponding fits, respectively. Black arrows mark the extracted phonon energies $E_{\rm ph}$ for the Stokes and anti-Stokes phonons. The vertical dashed lines are aligned to the phonon energies at the highest measured temperature. Temperature dependence of the fit values of $E_{\rm ph}$ in panels (a) and (b) are respectively shown in (c) and (d). The open and solid symbols are data from SPring-8 and APS, respectively. Vertical error bars are least-square fit errors of 1 s.d. (e) Schematic phase diagram of Sr$_{1-x}$Na$_x$Fe$_2$As$_2$, showing the antiferromagentic orthorhombic (AFM-O), antiferromagnetic tetragonal (AFM-T), paramagnetic tetragonal (PM-T), unidentified $C_2'$, and superconducting (SC) phases (adapted from Ref. \cite{liwang2019}). Red dots represent the two compositions studied in this work. Insets show the schematic magnetic structures for the AFM-O and AFM-T phases.}
\label{phonon}
\end{figure*}

For both the hole- and electron-doped sides of the FeSCs phase diagrams, the underdoped phase space is widely inhabited by intertwined nematic and magnetic orders \cite{Chu824,Chu2012,kim2011,rotundu2011,Kasahara2012}. Hole-doped FeSCs, such as Ba$_{1-x}$Na$_x$Fe$_2$As$_2$, Ba$_{1-x}$K$_x$Fe$_2$As$_2$ and Sr$_{1-x}$Na$_x$Fe$_2$As$_2$ \cite{Avci2014,branden2015,bohmer2015,Allred2016,taddei2016,taddei2017}, in addition uniquely exhibit a re-entrant tetragonal magnetic phase (AFM-T) within a small phase region in the underdoped regime, as well as several yet unidentified magnetic phases \cite{liwang2019,liwang2016}. Upon cooling, a system in this regime first transitions from a paramagnetic tetragonal (PM-T) phase into an orthorhombic phase with collinear antiferromagnetic order (AFM-O), where the striped magnetic order of the AFM-O phase (Fig.~\ref{phonon}(e)) couples to the nematic order with twofold rotational ($C_2$) symmetry. The system then transitions into a AFM-T phase which preserves fourfold rotational ($C_4$) symmetry (Fig.~\ref{phonon}(e))\cite{Allred2016}, and superconductivity eventually emerges as the ground state. 
The strength of the nematic fluctuations in a phase with $C_4$ symmetry can be characterized by the uniform (${\bf q}=0$) nematic susceptibility $\chi_{\rm nem}$. In the electron-doped side, $\chi_{\rm nem}$ has been probed using multiple experimental techniques, revealing a marked enhancement upon cooling \cite{fernandes2010,bohmer2016,yoshizawa2012,Kuo2016,kissikov2017,gallais2013,GALLAIS2016,WWang2018,Kretzschmar2016,sfwu2017}. On the hole-doped side, a similar enhancement is found in the PM-T phase \cite{bohmer2014}, and large nematic fluctuations persist in the AFM-T phase \cite{ben2017,LWang2018}.

An enhanced $\chi_{\rm nem}$ yields a softening of the shear modulus $C_{66}$ due to coupling with the lattice. This can be quantitatively studied through ultrasound \cite{fernandes2010,yoshizawa2012} or approximated by Young's modulus measurements \cite{bohmer2014,bohmer2016}. However, the spatial-dependence of the nematic susceptibility is not probed in these measurements. Whereas, due to coupling between the lattice and the nematic order parameter, the in-plane transverse acoustic (IPTA) phonons provide a measurement of both the uniform nematic susceptibility (the sound velocity when ${\bf q}\rightarrow 0$) and the nematic correlation length $\xi$ (encoded in the  dispersion when ${\bf q}\neq0$). 
While momentum-dependent softening of the IPTA phonon has been qualitatively observed in the parent and electron-doped FeSCs \cite{niedziela2011,parshall2015,liyu2018}, only very recently has it become possible to extract quantitative  information on the nematic correlation length \cite{weber2018,merritt2020}. So far, the behaviors of the IPTA phonon and evolution of the nematic correlation length remain unexplored for the hole-doped FeSCs, especially in the AFM-T phase absent in the electron-doped compounds.

The limitation in extracting $\xi$ in previous works \cite{weber2018, merritt2020} is the requirement of having as input both the bare shear modulus ($C_{66,0}$, without or with minimal nematic fluctuations \cite{yoshizawa2012}) and the renormalized shear modulus ($C_{66}$, renormalization due to coupling of the lattice with the nematic order parameter). However, on the one hand, $C_{66}$ is doping-dependent and has not been reported so far for Sr$_{1-x}$Na$_{x}$Fe$_2$As$_2$; on the other hand, an accurate measurement of $C_{66,0}$ is difficult since substantial nematic fluctuations exist in under- and optimal-doped samples up to high temperatures \cite{ben2018}. It is thus desirable to circumvent the above limitation and extract the nematic correlation length solely from IPTA measurements, without relying on measurements from other probes. 

In this Letter, we present inelastic X-ray scattering measurements of the IPTA phonons in Sr$_{1-x}$Na$_x$Fe$_2$As$_2$ for an underdoped sample (UD, $x=0.36$ with superconducting temperature $T_{\rm c}$ = 8.7~K, PM-T to AFM-O transition temperature $T_{\rm S}=107$~K and AFM-O to AFM-T magnetic transition temperature $T_{\rm r} = 68$~K ) and an optimally doped sample (OP, $x=0.55$ with $T_{\rm c}$ = 37~K) (Fig. \ref{phonon} (e)). Because the marginal dimensionality for structural phase transitions associated with soft transverse acoustic phonon modes is 2 (in the absence of disorder), the mean-field theory is exact \cite{cowley1976,nielsen1977,karahsanovic2016}. By utilizing this fact and based on previous methods \cite{weber2018,merritt2020}, we demonstrate a generic method for extracting the nematic susceptibility and correlation length simultaneously, without the requirement of additional input. 
Applying our method to the two Sr$_{1-x}$Na$_x$Fe$_2$As$_2$ samples reveals that nematic correlations in the hole-doped FeSCs are much shorter-ranged than in the electron-doped FeSCs, and suggest that they may be related to the more robust superconductivity observed in the former.

High quality single crystals were synthesized using a self-flux method, and have been characterized using magnetization and dilatometry measurements \cite{liwang2019}. Phonon measurements were carried at the BL35XU beamline at SPring-8 \cite{baron}, Japan, and the 30-ID beamline \cite{Toellner,Said} at the Advanced Photon Source (APS), Argonne National Laboratory. Incident photon energies were fixed to 21.7476~keV at BL35XU and 23.7~keV at 30-ID beamline. The energy resolution at BL35XU is 1.4 and 1.6 meV for two different spectrometer configurations. 
More experimental details are provided in the Supplemental Material \cite{SI}. We reference momentum transfer ${\bf Q}$ in reduced lattice units, using the tetragonal 2-Fe unit cell with in-plane lattice constants $\approx3.88$ and $\approx3.86$~${\rm \AA}$ for the UD and OP samples, respectively. Our measurements were carried out at ${\bf Q}=(2,k,0)$, dominated by the IPTA phonon.

Clear phonon softening can be observed for UD and OP samples upon cooling in Figs. \ref{phonon} (a) \& (b). This trend can be quantified in the temperature dependence of $E_{\rm ph}$ for the two samples (Figs. \ref{phonon} (c) \& (d)). 
For the UD sample, $E_{\rm ph}$ decreases upon cooling toward $T_{\rm S}$, which separates the PM-T and AFM-O phases, consistent with previous observations in the parent and electron-doped FeSCs \cite{niedziela2011,parshall2015,weber2018,liyu2018,merritt2020}. We note that in the AFM-O phase ($T_{\rm r}<T<T_{\rm S}$), $E_{\rm ph}$ cannot be uniquely determined due to twinning, so those data are not included in our analysis.  Intriguingly, in the AFM-T phase ($T_{\rm c}<T<T_{\rm r}$), despite the re-entrance to a tetragonal structure, the acoustic phonon along $(2,k,0)$ remains soft, indicating substantial retention of nematic fluctuations in the AFM-T phase. This persists in the superconducting phase. 
For the OP sample, in which both nematic and magnetic orders are absent, $E_{\rm ph}$ is slightly reduced upon cooling down to 25~K ($<T_{\rm c}$) from 300 K, indicating a less significant softening compared to the UD sample as expected. 

\begin{figure}
\includegraphics[width=1.\columnwidth,clip,angle =0]{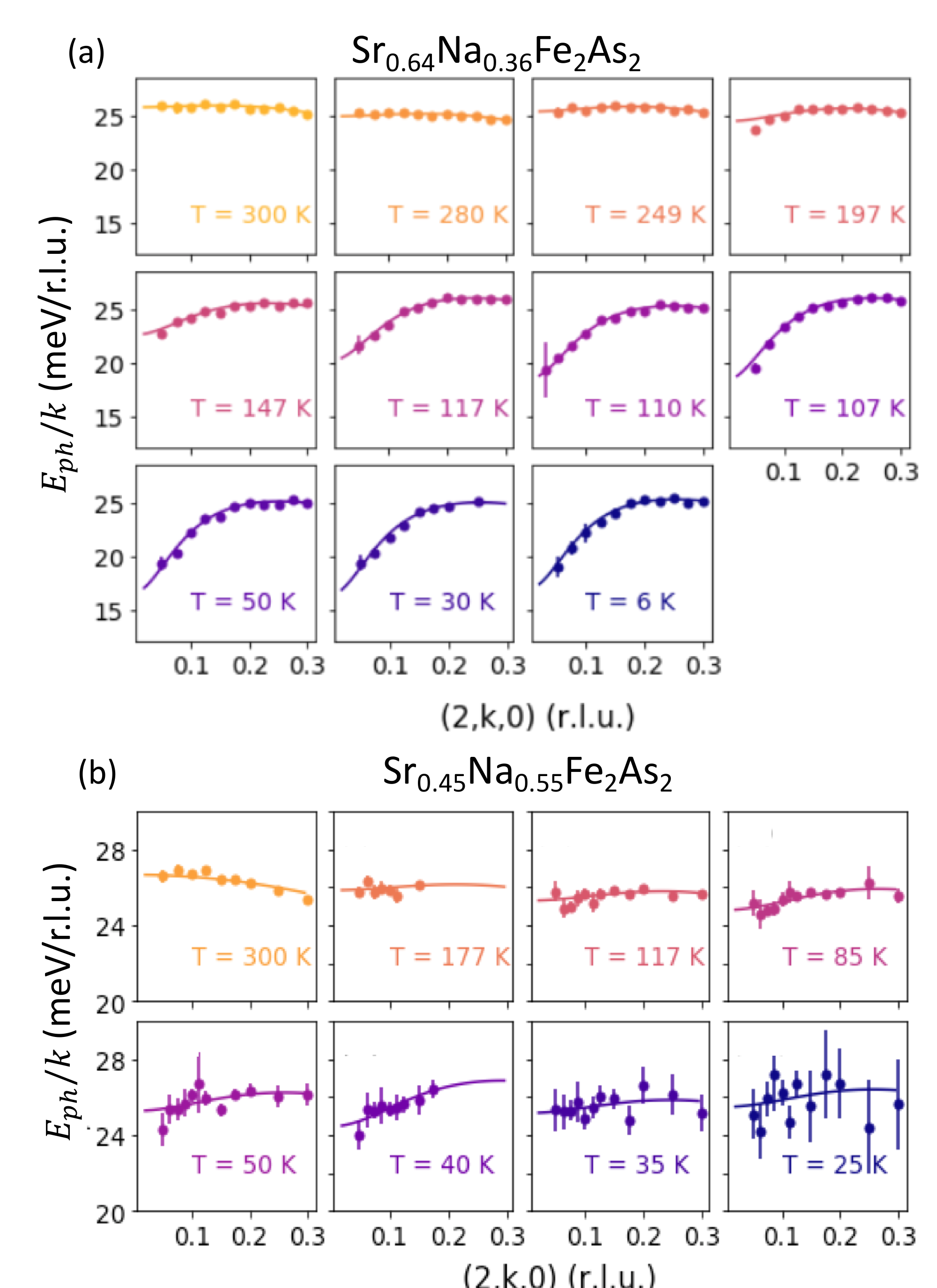}
\vspace{-1.5em}
\caption{\label{fit} $E_{\rm ph}/k$ plotted against $k$ (solid symbols) and corresponding fits (solid lines) for (a) Sr$_{0.64}$Na$_{0.36}$Fe$_2$As$_2$ and (b) Sr$_{0.45}$Na$_{0.55}$Fe$_2$As$_2$. A large nematic correlation length leads to more prominent phonon softening and nonlinearity in the plot of $E_{\rm ph}/k$ versus $k$. Details of the fits are described in the text. All vertical error bars are least-square fit errors of 1 s.d.}
\label{phononfit}
\end{figure}

To accentuate the abnormal phonon softening in momentum space, the quantity $E_{\rm ph}/k$ against $k$, called $k$-dependent sound velocity, is plotted as a function of temperature. For acoustic phonons without coupling to the electrons, the dispersion should be asymptotically linear in the long wavelength limit, therefore producing a constant $E_{\rm ph}/k$ over momentum (Fig. \ref{phononfit}). Phonon softening would, meanwhile, manifest as a deviation from the constant over a momentum range dictated by the nematic correlation length scale. Comparing the UD and OP sample, one immediate observation is that the range in which the nonlinearity develops is substantially larger in the UD ($k\lesssim0.2$) than the OP sample ($k\lesssim0.075$). 
By quantitatively analyzing the momentum- and temperature-dependence of these IPTA phonon softening using the mean-field theory, we obtain both the nematic susceptibility and correlation length without additional input of the shear modulus, as described below.

Energies of the IPTA phonon $E_{\rm ph}$ are renormalized due to coupling between the lattice and the nematic order parameter, with the spatial-dependence of the nematic correlations determining the momentum-dependence of the renormalizations. $E_{\rm ph}$ as a function of $k$ and the nematic correlation length $\xi$ are  related through \cite{weber2018}:
\begin{equation}
E_{\rm ph}(k) = f(k) \sqrt{\frac{ C_{66,0}}{\rho}} \sqrt {\frac{1+ \xi^2 k^2}{ \frac{C_{66,0}}{C_{66}} + \xi^2 k^2 }},
\label{eq1}
\end{equation}
\noindent where $\rho$ is the mass density.  The renormalized shear modulus $C_{66}$  is connected to the bare shear modulus $C_{66,0}$ through $\chi_{\rm nem}$ and the coupling strength $\lambda$ as
$C_{66,0}/C_{66} = 1 + \lambda^2 \chi_{\rm nem}/C_{66,0}$ \cite{fernandes2010}. $f(k)$ is the bare phonon dispersion \cite{merritt2020}.

The observation of mean-field behaviors for both the nematic susceptibility \cite{Chu2012} and nematic correlation length \cite{merritt2020} in FeSCs \cite{Kuo2016} suggests that nematic fluctuations obey mean-field behavior despite the Ising nature of the nematic order parameter. This results from long-range interactions mediated by the lattice. Specifically, the coupling between the Ising nematic order parameter and the lattice lowers the upper marginal dimension of the nematic order parameter from 4 to 2 thereby yielding classical mean-field behaviors in 3 dimensions \cite{cowley1976,nielsen1977,karahsanovic2016}. As a result, the nematic susceptibility and the nematic correlation length are predicted to exhibit mean-field behavior and are related through $\chi_{\rm nem}\propto\xi^2$. Using this relation, we obtain $\lambda^2 \chi_{\rm nem}/C_{66,0} = r \xi^2$, with $r$ being a proportionality variable independent of temperature. Eq. \ref{eq1} can then be rewritten as:

\begin{equation}
E_{\rm ph}(k) = f(k) \sqrt{\frac{ C_{66,0}}{\rho}} \sqrt {\frac{1+ \xi^2 k^2}{1+ \xi^2 (k^2 + r) }}.
\label{eq2}
\end{equation}
It should be noted that the above analysis ignores the effects of the random nematic fields created by the compositional disorder. We expect that the random fields will cause a crossover to non-mean field behavior at large enough length scales. However, our data almost certainly do not probe the random field critical regime so mean field theory should obtain.

In the UD sample, for $T> T_{\rm S}$ we have $\xi = \xi_0 (\frac{T}{T_0}-1)^{-\frac{1}{2}}$ when approaching a second-order phase transition \cite{nielsen1977,cowley1976}, further constraining our fitting. We expect that $T_0<T_{\rm S}$, as the phase transition between the PM-T phase and the AFM-O phase at $T_{\rm S}$ is, in fact, weakly first order \cite{liwang2019, taddei2016}.
Therefore, only four temperature-independent parameters - $C_{66,0}$, $\xi_0$, $r$ and $T_0$ - are sufficient in principle to describe the full data set. In our fitting procedure, we allowed for a temperature dependence of $C_{66,0}(T)$ and confirmed that it exhibits little or no variation with temperature or doping \cite{SI}.
In the AFM-T phase, we assumed that $\chi_{\rm nem}$ and $\xi$ are also related through $r$ as in the PM-T phase ($T>T_{\rm S}$), without additional constraints on $\xi$ . We have attempted to allow for different values of $r$ in the PM-T and AFM-T phases, and found the fitting results to be identical within fitting uncertainties \cite{SI}. For the OP sample, the data were fit with unconstrained $\xi$, as done for the AFM-T phase in the UD sample. 

\begin{figure}
\includegraphics[width=1.\columnwidth,clip,angle =0]{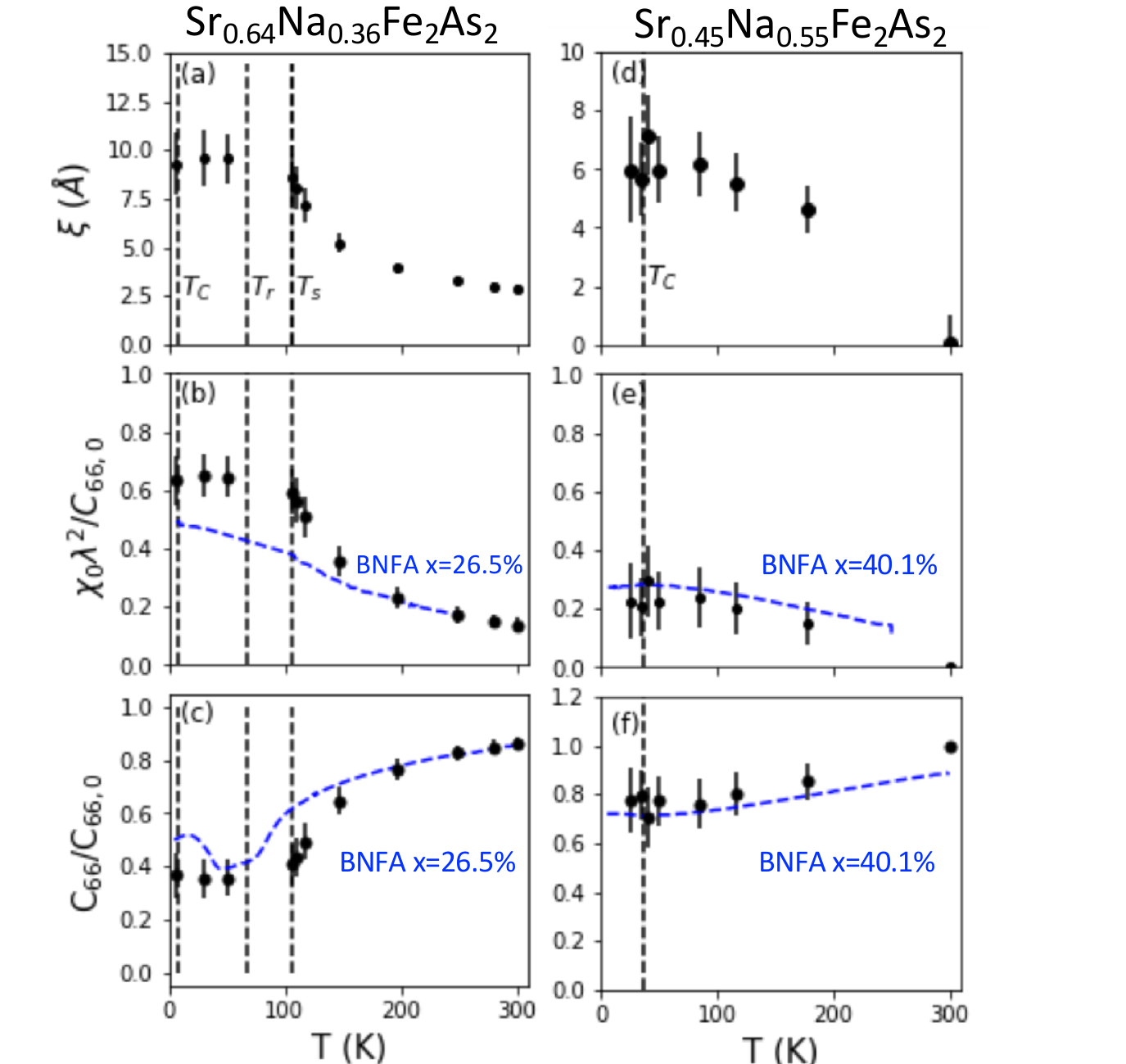}
\vspace{-1.5em}
\caption{\label{cl} Nematic correlation length $\xi$, derived bare nematic susceptibility in units of $\lambda^2/C_{66,0}$ and renormalized shear modulus $C_{66}/C_{66,0}$ as a function of temperature for (a)-(c) Sr$_{0.64}$Na$_{0.36}$Fe$_2$As$_2$ and (d)-(f) Sr$_{0.45}$Na$_{0.55}$Fe$_2$As$_2$.  
Here $\lambda$ is the coupling strength between the nematic order parameter and the lattice. $C_{66,0}$ is the bare shear modulus in the absence of nematic fluctuations. The vertical dashed lines mark phase transitions in the two samples. For the comparison only, the blue dashed lines are the bare nematic susceptibilities and the Young's moduli normalized to the unperturbed Young's modulus, $Y_{[110]}/Y_{[110]}^0$, in Ba$_{1-x}$Na$_x$Fe$_2$As$_2$ with $x=0.26$ (UD) and $x=0.4$ (OP), respectively\cite{LWang2018}. 
All vertical error bars are either least-square fit errors of 1 s.d., or obtained through the propagation of fit errors.
}
\label{result}
\end{figure}

The fits for both samples are shown in Fig. \ref{phononfit} as solid lines, and the temperature dependence of $\xi$ is plotted in Fig. \ref{result}. We find $\xi_0 = 4.3(6)$~${\rm \AA}$ in the UD sample, which is close to its in-plane lattice parameter. The extracted bare shear modulus $C_{66,0}$ is 39(2) GPa for the two samples, softer than the shear modulus of heavily electron-doped Ba(Fe$_{1-x}$Co$_x$)$_2$As$_2$ ($x=0.245$) where little nematic fluctuation exists \cite{yoshizawa2012}.
The fitted effective second order transition temperature $T_0= 90(5)$~K for the UD sample is somewhat below $T_{\rm S} = 107$~K, consistent with the notion that a weakly first order transition preempts a second order phase transition. 

The temperature-dependence of the bare nematic susceptibility and shear modulus derived from our fits can be  compared with Young's modulus measurements. 
Within a mean-field approach, due to coupling between the nematic order parameter and the lattice, the renormalized nematic susceptibility $\chi_{\rm nem} $ and the bare nematic susceptibility $\chi_0$ are related through the relation $\chi_{\rm nem}^{-1}  = \chi_0^{-1} -  \lambda^2/C_{66,0}$ \cite{fernandes2010}. Combined with $\chi_{\rm nem}=r\xi^2$, $\chi_0$ in units of $\lambda^2/C_{66,0}$ can be found through $\lambda^2 \chi_0/C_{66,0}=r\xi^2/(1+r\xi^2)$. The ratio $C_{66}/C_{66,0}$ can be found through $C_{66}/C_{66,0}=1-\lambda^2\chi_0/C_{66,0}$. We compare $\chi_0$ of UD and OP Sr$_{1-x}$Na$_x$Fe$_2$As$_2$ samples from our phonon measurements, with similarly doped Ba$_{1-x}$Na$_x$Fe$_2$As$_2$ samples from Young's modulus measurements \cite{LWang2018} in Figs. \ref{result} (b) \& (e). Qualitatively similar behaviors are found for the two series using different experimental techniques. The larger value in Sr$_{1-x}$Na$_x$Fe$_2$As$_2$ ($x=0.36$ ) is consistent with its larger $T_{\rm S}$ (106~K compared to $T_{\rm S}=79$~K in Ba$_{1-x}$Na$_x$Fe$_2$As$_2$ with $x=0.265$). Analogous comparisons are made between renormalized shear modulus $C_{66}/C_{66,0}$ and Young's modulus $Y_{[110]}/Y_{[110]}^0$ in Figs. \ref{result} (c) \& (f). The consistency of results from the two experimental techniques demonstrates that phonon measurements can be reliably utilized to obtain both the susceptibility and the correlation length of the nematic order parameter in the FeSCs.

Our findings reveal  important characteristics of the nematic fluctuations in the AFM-T phase.
Despite its tetragonal structure, our results reveal that the AFM-T phase exhibits a large nematic susceptibility, confirming the presence of significant nematic fluctuations \cite{RYu2017,LWang2018}. We further extract the nematic correlation length and show that both the nematic susceptibility and the nematic correlation length in the AFM-T phase remain similar to those just above $T_{\rm S}$. This suggests that the AFM-T phase acts to inhibit further development of nematic fluctuations. The value of $\xi\sim 10$ $\rm \AA$ in the AFM-T phase is consistent with the scale of local orthorhombic domains observed from pair distribution function analysis \cite{ben2017, ben2018} and indicate a fluctuating rather than pinned nature of these local orthorhombic domains.

Our results imply a role of the nematic correlation length in determining the system's electronic state, including the interaction with the superconductivity.
An asymmetry in the phase diagrams of electron- and hole-doped FeSCs is notable. Hole-doping results in a higher $T_{\rm c}$, a superconducting state that prevails for a much broader doping range \cite{PDai2015}, involvement of a AFM-T phase \cite{Avci2014} unique to hole-doped side, and possibly a superconducting state with broken time-reversal symmetry \cite{Grinenko2020}. Factors that may contribute to these unusual features include the dominance of $c$-axis polarized spin excitations at low energies ($\lesssim10$~meV) \cite{YSong2016,yuan2019}, an important role of spin-orbit coupling \cite{christensen2015}, random nematic field effects due to dopant-induced disorder effects \cite{timmons2019,hoyer2016,gastiasoro2015}, and the different strength of pair breaking in electron- and hole-doped iron-pnictides \cite{hardy2016}. 
Our work reveals a strong electron-hole asymmetry for $\xi_0$ in the UD regime of the FeSCs, with a much smaller $\xi_0$ on the hole-doped side ($\xi_0\sim 4$~{\AA}), compared to the electron-doped side ($\xi_0 \sim 40$~{\AA}). A similar contrast is seen for $\xi$ in the OD regime when approaching the SC phase, with $\xi\lesssim10$~{\AA} on the hole-doped side, and $\xi\sim100$~{\AA} on the electron-doped side \cite{SI}. 

Our findings demonstrate that, on an empirical level, a short nematic correlation length in the hole-doped FeSCs is correlated with a more robust superconducting dome and a higher optimal $T_{\rm c}$. This may be related to the fact that a short correlation length implies a weak tendency towards long-range nematic or magnetic order, which competes with superconductivity \cite{DPratt2009,SNandi2010,Yi2014,WWang2018,bohmer2015}.
While previous studies investigated how the nematic fluctuations could enhance superconducting pairing \cite{PhysRevLett.114.097001} or affect the pairing symmetry \cite{fernandes2013}, our observations call for experimental and theoretical works to examine the relation between the nematic correlation length and the superconductivity.

More broadly, our results suggest the correlation lengths of various electronic order parameters should be investigated in the FeSCs to understand their role in the superconductivity.
Specific topics that warrant careful examination include comparison of the magnetic \cite{SWilson2010} and nematic correlation lengths, with the superconducting coherence length \cite{MAltarawneh2008,Yuan2009}.
In addition, random nematic field effects \cite{YLoh2010,Carlson2011} due to compositional disorder should also be studied more thoroughly with regards to their impacts on limiting the nematic and magnetic correlation lengths, as well as how quantum criticality is in turn affected \cite{zhang2019,Kuo2016}. 
Finally, the method of analysis that we have introduced here should be applicable to other systems exhibiting coupled nematic-soft transverse acoustic phonon phase transitions, in which in the absence of disorder, the nematic order parameter follows mean-field behavior exactly due to coupling with the lattice.

This work is funded by the U.S. Department of Energy, Office of Science, Office of Basic Energy Sciences, Materials Sciences and Engineering Division under Contract No. DE-AC02-05-CH11231 within the Quantum Materials Program (KC2202). AF acknowledges support from the Alfred P. Sloan Foundation. YH acknowledges support from the Miller Institute for Basic Research in Science. MY acknowledges support from the Alfred P. Sloan Foundation as well as the Robert A. Welch Foundation Grant No. C-2024. Measurements at the BL35XU beamline at SPring-8 under approval from JASRI were performed under Proposal No. 2019B1455.  This research used resources of the Advanced Photon Source, a U.S. Department of Energy (DOE) Office of Science User Facility operated for the DOE Office of Science by Argonne National Laboratory under Contract No. DE-AC02-06CH11357.

\bibliography{bibfile}

\begin{thebibliography}{63}
\expandafter\ifx\csname natexlab\endcsname\relax\def\natexlab#1{#1}\fi
\expandafter\ifx\csname bibnamefont\endcsname\relax
  \def\bibnamefont#1{#1}\fi
\expandafter\ifx\csname bibfnamefont\endcsname\relax
  \def\bibfnamefont#1{#1}\fi
\expandafter\ifx\csname citenamefont\endcsname\relax
  \def\citenamefont#1{#1}\fi
\expandafter\ifx\csname url\endcsname\relax
  \def\url#1{\texttt{#1}}\fi
\expandafter\ifx\csname urlprefix\endcsname\relax\def\urlprefix{URL }\fi
\providecommand{\bibinfo}[2]{#2}
\providecommand{\eprint}[2][]{\url{#2}}

\bibitem[{\citenamefont{Fernandes and Schmalian}(2012)}]{Fernandes_2012}
\bibinfo{author}{\bibfnamefont{R.~M.} \bibnamefont{Fernandes}}
  \bibnamefont{and}
  \bibinfo{author}{\bibfnamefont{J.}~\bibnamefont{Schmalian}},
  \bibinfo{journal}{Superconductor Science and Technology}
  \textbf{\bibinfo{volume}{25}}, \bibinfo{pages}{084005}
  (\bibinfo{year}{2012}).

\bibitem[{\citenamefont{Kuo et~al.}(2016)\citenamefont{Kuo, Chu, Palmstrom,
  Kivelson, and Fisher}}]{Kuo2016}
\bibinfo{author}{\bibfnamefont{H.-H.} \bibnamefont{Kuo}},
  \bibinfo{author}{\bibfnamefont{J.-H.} \bibnamefont{Chu}},
  \bibinfo{author}{\bibfnamefont{J.~C.} \bibnamefont{Palmstrom}},
  \bibinfo{author}{\bibfnamefont{S.~A.} \bibnamefont{Kivelson}},
  \bibnamefont{and} \bibinfo{author}{\bibfnamefont{I.~R.}
  \bibnamefont{Fisher}}, \bibinfo{journal}{Science}
  \textbf{\bibinfo{volume}{352}}, \bibinfo{pages}{958} (\bibinfo{year}{2016}).

\bibitem[{\citenamefont{Fradkin et~al.}(2010)\citenamefont{Fradkin, Kivelson,
  Lawler, Eisenstein, and Mackenzie}}]{fradkin2010}
\bibinfo{author}{\bibfnamefont{E.}~\bibnamefont{Fradkin}},
  \bibinfo{author}{\bibfnamefont{S.~A.} \bibnamefont{Kivelson}},
  \bibinfo{author}{\bibfnamefont{M.~J.} \bibnamefont{Lawler}},
  \bibinfo{author}{\bibfnamefont{J.~P.} \bibnamefont{Eisenstein}},
  \bibnamefont{and} \bibinfo{author}{\bibfnamefont{A.~P.}
  \bibnamefont{Mackenzie}}, \bibinfo{journal}{Annual Review of Condensed Matter
  Physics} \textbf{\bibinfo{volume}{1}}, \bibinfo{pages}{153}
  (\bibinfo{year}{2010}).

\bibitem[{\citenamefont{Fernandes et~al.}(2014)\citenamefont{Fernandes,
  Chubukov, and Schmalian}}]{fernandes2014}
\bibinfo{author}{\bibfnamefont{R.~M.} \bibnamefont{Fernandes}},
  \bibinfo{author}{\bibfnamefont{A.~V.} \bibnamefont{Chubukov}},
  \bibnamefont{and}
  \bibinfo{author}{\bibfnamefont{J.}~\bibnamefont{Schmalian}},
  \bibinfo{journal}{Nature Physics} \textbf{\bibinfo{volume}{10}},
  \bibinfo{pages}{97} (\bibinfo{year}{2014}), ISSN \bibinfo{issn}{1745-2481}.

\bibitem[{\citenamefont{Chu et~al.}(2012)\citenamefont{Chu, Kuo, Analytis, and
  Fisher}}]{Chu2012}
\bibinfo{author}{\bibfnamefont{J.-H.} \bibnamefont{Chu}},
  \bibinfo{author}{\bibfnamefont{H.-H.} \bibnamefont{Kuo}},
  \bibinfo{author}{\bibfnamefont{J.~G.} \bibnamefont{Analytis}},
  \bibnamefont{and} \bibinfo{author}{\bibfnamefont{I.~R.}
  \bibnamefont{Fisher}}, \bibinfo{journal}{Science}
  \textbf{\bibinfo{volume}{337}}, \bibinfo{pages}{710} (\bibinfo{year}{2012}).

\bibitem[{\citenamefont{Chu et~al.}(2010)\citenamefont{Chu, Analytis, De~Greve,
  McMahon, Islam, Yamamoto, and Fisher}}]{Chu824}
\bibinfo{author}{\bibfnamefont{J.-H.} \bibnamefont{Chu}},
  \bibinfo{author}{\bibfnamefont{J.~G.} \bibnamefont{Analytis}},
  \bibinfo{author}{\bibfnamefont{K.}~\bibnamefont{De~Greve}},
  \bibinfo{author}{\bibfnamefont{P.~L.} \bibnamefont{McMahon}},
  \bibinfo{author}{\bibfnamefont{Z.}~\bibnamefont{Islam}},
  \bibinfo{author}{\bibfnamefont{Y.}~\bibnamefont{Yamamoto}}, \bibnamefont{and}
  \bibinfo{author}{\bibfnamefont{I.~R.} \bibnamefont{Fisher}},
  \bibinfo{journal}{Science} \textbf{\bibinfo{volume}{329}},
  \bibinfo{pages}{824} (\bibinfo{year}{2010}), ISSN \bibinfo{issn}{0036-8075},
  \eprint{https://science.sciencemag.org/content/329/5993/824.full.pdf}.

\bibitem[{\citenamefont{Kim et~al.}(2011)\citenamefont{Kim, Fernandes,
  Kreyssig, Kim, Thaler, Bud'ko, Canfield, McQueeney, Schmalian, and
  Goldman}}]{kim2011}
\bibinfo{author}{\bibfnamefont{M.~G.} \bibnamefont{Kim}},
  \bibinfo{author}{\bibfnamefont{R.~M.} \bibnamefont{Fernandes}},
  \bibinfo{author}{\bibfnamefont{A.}~\bibnamefont{Kreyssig}},
  \bibinfo{author}{\bibfnamefont{J.~W.} \bibnamefont{Kim}},
  \bibinfo{author}{\bibfnamefont{A.}~\bibnamefont{Thaler}},
  \bibinfo{author}{\bibfnamefont{S.~L.} \bibnamefont{Bud'ko}},
  \bibinfo{author}{\bibfnamefont{P.~C.} \bibnamefont{Canfield}},
  \bibinfo{author}{\bibfnamefont{R.~J.} \bibnamefont{McQueeney}},
  \bibinfo{author}{\bibfnamefont{J.}~\bibnamefont{Schmalian}},
  \bibnamefont{and} \bibinfo{author}{\bibfnamefont{A.~I.}
  \bibnamefont{Goldman}}, \bibinfo{journal}{Phys. Rev. B}
  \textbf{\bibinfo{volume}{83}}, \bibinfo{pages}{134522}
  (\bibinfo{year}{2011}).

\bibitem[{\citenamefont{Rotundu and Birgeneau}(2011)}]{rotundu2011}
\bibinfo{author}{\bibfnamefont{C.~R.} \bibnamefont{Rotundu}} \bibnamefont{and}
  \bibinfo{author}{\bibfnamefont{R.~J.} \bibnamefont{Birgeneau}},
  \bibinfo{journal}{Phys. Rev. B} \textbf{\bibinfo{volume}{84}},
  \bibinfo{pages}{092501} (\bibinfo{year}{2011}).

\bibitem[{\citenamefont{Kasahara et~al.}(2012)\citenamefont{Kasahara, Shi,
  Hashimoto, Tonegawa, Mizukami, Shibauchi, Sugimoto, Fukuda, Terashima,
  Nevidomskyy et~al.}}]{Kasahara2012}
\bibinfo{author}{\bibfnamefont{S.}~\bibnamefont{Kasahara}},
  \bibinfo{author}{\bibfnamefont{H.~J.} \bibnamefont{Shi}},
  \bibinfo{author}{\bibfnamefont{K.}~\bibnamefont{Hashimoto}},
  \bibinfo{author}{\bibfnamefont{S.}~\bibnamefont{Tonegawa}},
  \bibinfo{author}{\bibfnamefont{Y.}~\bibnamefont{Mizukami}},
  \bibinfo{author}{\bibfnamefont{T.}~\bibnamefont{Shibauchi}},
  \bibinfo{author}{\bibfnamefont{K.}~\bibnamefont{Sugimoto}},
  \bibinfo{author}{\bibfnamefont{T.}~\bibnamefont{Fukuda}},
  \bibinfo{author}{\bibfnamefont{T.}~\bibnamefont{Terashima}},
  \bibinfo{author}{\bibfnamefont{A.~H.} \bibnamefont{Nevidomskyy}},
  \bibnamefont{et~al.}, \bibinfo{journal}{Nature}
  \textbf{\bibinfo{volume}{486}}, \bibinfo{pages}{382} (\bibinfo{year}{2012}),
  ISSN \bibinfo{issn}{1476-4687}.

\bibitem[{\citenamefont{Avci et~al.}(2014)\citenamefont{Avci, Chmaissem,
  Allred, Rosenkranz, Eremin, Chubukov, Bugaris, Chung, Kanatzidis, Castellan
  et~al.}}]{Avci2014}
\bibinfo{author}{\bibfnamefont{S.}~\bibnamefont{Avci}},
  \bibinfo{author}{\bibfnamefont{O.}~\bibnamefont{Chmaissem}},
  \bibinfo{author}{\bibfnamefont{J.~M.} \bibnamefont{Allred}},
  \bibinfo{author}{\bibfnamefont{S.}~\bibnamefont{Rosenkranz}},
  \bibinfo{author}{\bibfnamefont{I.}~\bibnamefont{Eremin}},
  \bibinfo{author}{\bibfnamefont{A.~V.} \bibnamefont{Chubukov}},
  \bibinfo{author}{\bibfnamefont{D.~E.} \bibnamefont{Bugaris}},
  \bibinfo{author}{\bibfnamefont{D.~Y.} \bibnamefont{Chung}},
  \bibinfo{author}{\bibfnamefont{M.~G.} \bibnamefont{Kanatzidis}},
  \bibinfo{author}{\bibfnamefont{J.-P.} \bibnamefont{Castellan}},
  \bibnamefont{et~al.}, \bibinfo{journal}{Nature Communications}
  \textbf{\bibinfo{volume}{5}}, \bibinfo{pages}{3845} (\bibinfo{year}{2014}),
  ISSN \bibinfo{issn}{2041-1723}.

\bibitem[{\citenamefont{Wa\ss{}er et~al.}(2015)\citenamefont{Wa\ss{}er,
  Schneidewind, Sidis, Wurmehl, Aswartham, B\"uchner, and
  Braden}}]{branden2015}
\bibinfo{author}{\bibfnamefont{F.}~\bibnamefont{Wa\ss{}er}},
  \bibinfo{author}{\bibfnamefont{A.}~\bibnamefont{Schneidewind}},
  \bibinfo{author}{\bibfnamefont{Y.}~\bibnamefont{Sidis}},
  \bibinfo{author}{\bibfnamefont{S.}~\bibnamefont{Wurmehl}},
  \bibinfo{author}{\bibfnamefont{S.}~\bibnamefont{Aswartham}},
  \bibinfo{author}{\bibfnamefont{B.}~\bibnamefont{B\"uchner}},
  \bibnamefont{and} \bibinfo{author}{\bibfnamefont{M.}~\bibnamefont{Braden}},
  \bibinfo{journal}{Phys. Rev. B} \textbf{\bibinfo{volume}{91}},
  \bibinfo{pages}{060505} (\bibinfo{year}{2015}).

\bibitem[{\citenamefont{B{\"o}hmer et~al.}(2015)\citenamefont{B{\"o}hmer,
  Hardy, Wang, Wolf, Schweiss, and Meingast}}]{bohmer2015}
\bibinfo{author}{\bibfnamefont{A.~E.} \bibnamefont{B{\"o}hmer}},
  \bibinfo{author}{\bibfnamefont{F.}~\bibnamefont{Hardy}},
  \bibinfo{author}{\bibfnamefont{L.}~\bibnamefont{Wang}},
  \bibinfo{author}{\bibfnamefont{T.}~\bibnamefont{Wolf}},
  \bibinfo{author}{\bibfnamefont{P.}~\bibnamefont{Schweiss}}, \bibnamefont{and}
  \bibinfo{author}{\bibfnamefont{C.}~\bibnamefont{Meingast}},
  \bibinfo{journal}{Nature Communications} \textbf{\bibinfo{volume}{6}},
  \bibinfo{pages}{7911} (\bibinfo{year}{2015}).

\bibitem[{\citenamefont{Allred et~al.}(2016)\citenamefont{Allred, Taddei,
  Bugaris, Krogstad, Lapidus, Chung, Claus, Kanatzidis, Brown, Kang
  et~al.}}]{Allred2016}
\bibinfo{author}{\bibfnamefont{J.~M.} \bibnamefont{Allred}},
  \bibinfo{author}{\bibfnamefont{K.~M.} \bibnamefont{Taddei}},
  \bibinfo{author}{\bibfnamefont{D.~E.} \bibnamefont{Bugaris}},
  \bibinfo{author}{\bibfnamefont{M.~J.} \bibnamefont{Krogstad}},
  \bibinfo{author}{\bibfnamefont{S.~H.} \bibnamefont{Lapidus}},
  \bibinfo{author}{\bibfnamefont{D.~Y.} \bibnamefont{Chung}},
  \bibinfo{author}{\bibfnamefont{H.}~\bibnamefont{Claus}},
  \bibinfo{author}{\bibfnamefont{M.~G.} \bibnamefont{Kanatzidis}},
  \bibinfo{author}{\bibfnamefont{D.~E.} \bibnamefont{Brown}},
  \bibinfo{author}{\bibfnamefont{J.}~\bibnamefont{Kang}}, \bibnamefont{et~al.},
  \bibinfo{journal}{Nature Physics} \textbf{\bibinfo{volume}{12}},
  \bibinfo{pages}{493} (\bibinfo{year}{2016}), ISSN \bibinfo{issn}{1745-2481}.

\bibitem[{\citenamefont{Taddei et~al.}(2016)\citenamefont{Taddei, Allred,
  Bugaris, Lapidus, Krogstad, Stadel, Claus, Chung, Kanatzidis, Rosenkranz
  et~al.}}]{taddei2016}
\bibinfo{author}{\bibfnamefont{K.~M.} \bibnamefont{Taddei}},
  \bibinfo{author}{\bibfnamefont{J.~M.} \bibnamefont{Allred}},
  \bibinfo{author}{\bibfnamefont{D.~E.} \bibnamefont{Bugaris}},
  \bibinfo{author}{\bibfnamefont{S.}~\bibnamefont{Lapidus}},
  \bibinfo{author}{\bibfnamefont{M.~J.} \bibnamefont{Krogstad}},
  \bibinfo{author}{\bibfnamefont{R.}~\bibnamefont{Stadel}},
  \bibinfo{author}{\bibfnamefont{H.}~\bibnamefont{Claus}},
  \bibinfo{author}{\bibfnamefont{D.~Y.} \bibnamefont{Chung}},
  \bibinfo{author}{\bibfnamefont{M.~G.} \bibnamefont{Kanatzidis}},
  \bibinfo{author}{\bibfnamefont{S.}~\bibnamefont{Rosenkranz}},
  \bibnamefont{et~al.}, \bibinfo{journal}{Phys. Rev. B}
  \textbf{\bibinfo{volume}{93}}, \bibinfo{pages}{134510}
  (\bibinfo{year}{2016}).

\bibitem[{\citenamefont{Taddei et~al.}(2017)\citenamefont{Taddei, Allred,
  Bugaris, Lapidus, Krogstad, Claus, Chung, Kanatzidis, Osborn, Rosenkranz
  et~al.}}]{taddei2017}
\bibinfo{author}{\bibfnamefont{K.~M.} \bibnamefont{Taddei}},
  \bibinfo{author}{\bibfnamefont{J.~M.} \bibnamefont{Allred}},
  \bibinfo{author}{\bibfnamefont{D.~E.} \bibnamefont{Bugaris}},
  \bibinfo{author}{\bibfnamefont{S.~H.} \bibnamefont{Lapidus}},
  \bibinfo{author}{\bibfnamefont{M.~J.} \bibnamefont{Krogstad}},
  \bibinfo{author}{\bibfnamefont{H.}~\bibnamefont{Claus}},
  \bibinfo{author}{\bibfnamefont{D.~Y.} \bibnamefont{Chung}},
  \bibinfo{author}{\bibfnamefont{M.~G.} \bibnamefont{Kanatzidis}},
  \bibinfo{author}{\bibfnamefont{R.}~\bibnamefont{Osborn}},
  \bibinfo{author}{\bibfnamefont{S.}~\bibnamefont{Rosenkranz}},
  \bibnamefont{et~al.}, \bibinfo{journal}{Phys. Rev. B}
  \textbf{\bibinfo{volume}{95}}, \bibinfo{pages}{064508}
  (\bibinfo{year}{2017}).

\bibitem[{\citenamefont{Wang et~al.}(2019)\citenamefont{Wang, He, Scherer,
  Hardy, Schweiss, Wolf, Merz, Andersen, and Meingast}}]{liwang2019}
\bibinfo{author}{\bibfnamefont{L.}~\bibnamefont{Wang}},
  \bibinfo{author}{\bibfnamefont{M.}~\bibnamefont{He}},
  \bibinfo{author}{\bibfnamefont{D.~D.} \bibnamefont{Scherer}},
  \bibinfo{author}{\bibfnamefont{F.}~\bibnamefont{Hardy}},
  \bibinfo{author}{\bibfnamefont{P.}~\bibnamefont{Schweiss}},
  \bibinfo{author}{\bibfnamefont{T.}~\bibnamefont{Wolf}},
  \bibinfo{author}{\bibfnamefont{M.}~\bibnamefont{Merz}},
  \bibinfo{author}{\bibfnamefont{B.~M.} \bibnamefont{Andersen}},
  \bibnamefont{and} \bibinfo{author}{\bibfnamefont{C.}~\bibnamefont{Meingast}},
  \bibinfo{journal}{Journal of the Physical Society of Japan}
  \textbf{\bibinfo{volume}{88}}, \bibinfo{pages}{104710}
  (\bibinfo{year}{2019}).

\bibitem[{\citenamefont{Wang et~al.}(2016)\citenamefont{Wang, Hardy, B\"ohmer,
  Wolf, Schweiss, and Meingast}}]{liwang2016}
\bibinfo{author}{\bibfnamefont{L.}~\bibnamefont{Wang}},
  \bibinfo{author}{\bibfnamefont{F.}~\bibnamefont{Hardy}},
  \bibinfo{author}{\bibfnamefont{A.~E.} \bibnamefont{B\"ohmer}},
  \bibinfo{author}{\bibfnamefont{T.}~\bibnamefont{Wolf}},
  \bibinfo{author}{\bibfnamefont{P.}~\bibnamefont{Schweiss}}, \bibnamefont{and}
  \bibinfo{author}{\bibfnamefont{C.}~\bibnamefont{Meingast}},
  \bibinfo{journal}{Phys. Rev. B} \textbf{\bibinfo{volume}{93}},
  \bibinfo{pages}{014514} (\bibinfo{year}{2016}).

\bibitem[{\citenamefont{Fernandes et~al.}(2010)\citenamefont{Fernandes,
  VanBebber, Bhattacharya, Chandra, Keppens, Mandrus, McGuire, Sales, Sefat,
  and Schmalian}}]{fernandes2010}
\bibinfo{author}{\bibfnamefont{R.~M.} \bibnamefont{Fernandes}},
  \bibinfo{author}{\bibfnamefont{L.~H.} \bibnamefont{VanBebber}},
  \bibinfo{author}{\bibfnamefont{S.}~\bibnamefont{Bhattacharya}},
  \bibinfo{author}{\bibfnamefont{P.}~\bibnamefont{Chandra}},
  \bibinfo{author}{\bibfnamefont{V.}~\bibnamefont{Keppens}},
  \bibinfo{author}{\bibfnamefont{D.}~\bibnamefont{Mandrus}},
  \bibinfo{author}{\bibfnamefont{M.~A.} \bibnamefont{McGuire}},
  \bibinfo{author}{\bibfnamefont{B.~C.} \bibnamefont{Sales}},
  \bibinfo{author}{\bibfnamefont{A.~S.} \bibnamefont{Sefat}}, \bibnamefont{and}
  \bibinfo{author}{\bibfnamefont{J.}~\bibnamefont{Schmalian}},
  \bibinfo{journal}{Phys. Rev. Lett.} \textbf{\bibinfo{volume}{105}},
  \bibinfo{pages}{157003} (\bibinfo{year}{2010}).

\bibitem[{\citenamefont{B{\"o}hmer and Meingast}(2016)}]{bohmer2016}
\bibinfo{author}{\bibfnamefont{A.~E.} \bibnamefont{B{\"o}hmer}}
  \bibnamefont{and} \bibinfo{author}{\bibfnamefont{C.}~\bibnamefont{Meingast}},
  \bibinfo{journal}{Comptes Rendus Physique} \textbf{\bibinfo{volume}{17}},
  \bibinfo{pages}{90 } (\bibinfo{year}{2016}), ISSN \bibinfo{issn}{1631-0705}.

\bibitem[{\citenamefont{Yoshizawa et~al.}(2012)\citenamefont{Yoshizawa, Kimura,
  Chiba, Simayi, Nakanishi, Kihou, Lee, Iyo, Eisaki, Nakajima
  et~al.}}]{yoshizawa2012}
\bibinfo{author}{\bibfnamefont{M.}~\bibnamefont{Yoshizawa}},
  \bibinfo{author}{\bibfnamefont{D.}~\bibnamefont{Kimura}},
  \bibinfo{author}{\bibfnamefont{T.}~\bibnamefont{Chiba}},
  \bibinfo{author}{\bibfnamefont{S.}~\bibnamefont{Simayi}},
  \bibinfo{author}{\bibfnamefont{Y.}~\bibnamefont{Nakanishi}},
  \bibinfo{author}{\bibfnamefont{K.}~\bibnamefont{Kihou}},
  \bibinfo{author}{\bibfnamefont{C.-H.} \bibnamefont{Lee}},
  \bibinfo{author}{\bibfnamefont{A.}~\bibnamefont{Iyo}},
  \bibinfo{author}{\bibfnamefont{H.}~\bibnamefont{Eisaki}},
  \bibinfo{author}{\bibfnamefont{M.}~\bibnamefont{Nakajima}},
  \bibnamefont{et~al.}, \bibinfo{journal}{Journal of the Physical Society of
  Japan} \textbf{\bibinfo{volume}{81}}, \bibinfo{pages}{024604}
  (\bibinfo{year}{2012}).

\bibitem[{\citenamefont{Kissikov et~al.}(2017)\citenamefont{Kissikov, Sarkar,
  Lawson, Bush, Timmons, Tanatar, Prozorov, Bud'ko, Canfield, Fernandes
  et~al.}}]{kissikov2017}
\bibinfo{author}{\bibfnamefont{T.}~\bibnamefont{Kissikov}},
  \bibinfo{author}{\bibfnamefont{R.}~\bibnamefont{Sarkar}},
  \bibinfo{author}{\bibfnamefont{M.}~\bibnamefont{Lawson}},
  \bibinfo{author}{\bibfnamefont{B.~T.} \bibnamefont{Bush}},
  \bibinfo{author}{\bibfnamefont{E.~I.} \bibnamefont{Timmons}},
  \bibinfo{author}{\bibfnamefont{M.~A.} \bibnamefont{Tanatar}},
  \bibinfo{author}{\bibfnamefont{R.}~\bibnamefont{Prozorov}},
  \bibinfo{author}{\bibfnamefont{S.~L.} \bibnamefont{Bud'ko}},
  \bibinfo{author}{\bibfnamefont{P.~C.} \bibnamefont{Canfield}},
  \bibinfo{author}{\bibfnamefont{R.~M.} \bibnamefont{Fernandes}},
  \bibnamefont{et~al.}, \bibinfo{journal}{Phys. Rev. B}
  \textbf{\bibinfo{volume}{96}}, \bibinfo{pages}{241108}
  (\bibinfo{year}{2017}).

\bibitem[{\citenamefont{Gallais et~al.}(2013)\citenamefont{Gallais, Fernandes,
  Paul, Chauvi\`ere, Yang, M\'easson, Cazayous, Sacuto, Colson, and
  Forget}}]{gallais2013}
\bibinfo{author}{\bibfnamefont{Y.}~\bibnamefont{Gallais}},
  \bibinfo{author}{\bibfnamefont{R.~M.} \bibnamefont{Fernandes}},
  \bibinfo{author}{\bibfnamefont{I.}~\bibnamefont{Paul}},
  \bibinfo{author}{\bibfnamefont{L.}~\bibnamefont{Chauvi\`ere}},
  \bibinfo{author}{\bibfnamefont{Y.-X.} \bibnamefont{Yang}},
  \bibinfo{author}{\bibfnamefont{M.-A.} \bibnamefont{M\'easson}},
  \bibinfo{author}{\bibfnamefont{M.}~\bibnamefont{Cazayous}},
  \bibinfo{author}{\bibfnamefont{A.}~\bibnamefont{Sacuto}},
  \bibinfo{author}{\bibfnamefont{D.}~\bibnamefont{Colson}}, \bibnamefont{and}
  \bibinfo{author}{\bibfnamefont{A.}~\bibnamefont{Forget}},
  \bibinfo{journal}{Phys. Rev. Lett.} \textbf{\bibinfo{volume}{111}},
  \bibinfo{pages}{267001} (\bibinfo{year}{2013}).

\bibitem[{\citenamefont{Gallais and Paul}(2016)}]{GALLAIS2016}
\bibinfo{author}{\bibfnamefont{Y.}~\bibnamefont{Gallais}} \bibnamefont{and}
  \bibinfo{author}{\bibfnamefont{I.}~\bibnamefont{Paul}},
  \bibinfo{journal}{Comptes Rendus Physique} \textbf{\bibinfo{volume}{17}},
  \bibinfo{pages}{113 } (\bibinfo{year}{2016}), ISSN \bibinfo{issn}{1631-0705},
  \bibinfo{note}{iron-based superconductors / Supraconducteurs à base de fer}.

\bibitem[{\citenamefont{Wang et~al.}(2018{\natexlab{a}})\citenamefont{Wang,
  Song, Cao, Tseng, Keller, Li, Harriger, Tian, Chi, Yu et~al.}}]{WWang2018}
\bibinfo{author}{\bibfnamefont{W.}~\bibnamefont{Wang}},
  \bibinfo{author}{\bibfnamefont{Y.}~\bibnamefont{Song}},
  \bibinfo{author}{\bibfnamefont{C.}~\bibnamefont{Cao}},
  \bibinfo{author}{\bibfnamefont{K.-F.} \bibnamefont{Tseng}},
  \bibinfo{author}{\bibfnamefont{T.}~\bibnamefont{Keller}},
  \bibinfo{author}{\bibfnamefont{Y.}~\bibnamefont{Li}},
  \bibinfo{author}{\bibfnamefont{L.~W.} \bibnamefont{Harriger}},
  \bibinfo{author}{\bibfnamefont{W.}~\bibnamefont{Tian}},
  \bibinfo{author}{\bibfnamefont{S.}~\bibnamefont{Chi}},
  \bibinfo{author}{\bibfnamefont{R.}~\bibnamefont{Yu}}, \bibnamefont{et~al.},
  \bibinfo{journal}{Nature Communications} \textbf{\bibinfo{volume}{9}}
  (\bibinfo{year}{2018}{\natexlab{a}}).

\bibitem[{\citenamefont{Kretzschmar et~al.}(2016)\citenamefont{Kretzschmar,
  B\"{o}hm, Karahasanovi\'{c}, Muschler, Baum, Jost, Schmalian, Caprara,
  Grilli, Castro et~al.}}]{Kretzschmar2016}
\bibinfo{author}{\bibfnamefont{F.}~\bibnamefont{Kretzschmar}},
  \bibinfo{author}{\bibfnamefont{T.}~\bibnamefont{B\"{o}hm}},
  \bibinfo{author}{\bibfnamefont{U.}~\bibnamefont{Karahasanovi\'{c}}},
  \bibinfo{author}{\bibfnamefont{B.}~\bibnamefont{Muschler}},
  \bibinfo{author}{\bibfnamefont{A.}~\bibnamefont{Baum}},
  \bibinfo{author}{\bibfnamefont{D.}~\bibnamefont{Jost}},
  \bibinfo{author}{\bibfnamefont{J.}~\bibnamefont{Schmalian}},
  \bibinfo{author}{\bibfnamefont{S.}~\bibnamefont{Caprara}},
  \bibinfo{author}{\bibfnamefont{M.}~\bibnamefont{Grilli}},
  \bibinfo{author}{\bibfnamefont{C.~D.} \bibnamefont{Castro}},
  \bibnamefont{et~al.}, \bibinfo{journal}{Nature Physics}
  \textbf{\bibinfo{volume}{12}}, \bibinfo{pages}{560} (\bibinfo{year}{2016}),
  \urlprefix\url{https://doi.org/10.1038/nphys3634}.

\bibitem[{\citenamefont{Wu et~al.}(2017)\citenamefont{Wu, Richard, Ding, Wen,
  Tan, Wang, Zhang, Dai, and Blumberg}}]{sfwu2017}
\bibinfo{author}{\bibfnamefont{S.-F.} \bibnamefont{Wu}},
  \bibinfo{author}{\bibfnamefont{P.}~\bibnamefont{Richard}},
  \bibinfo{author}{\bibfnamefont{H.}~\bibnamefont{Ding}},
  \bibinfo{author}{\bibfnamefont{H.-H.} \bibnamefont{Wen}},
  \bibinfo{author}{\bibfnamefont{G.}~\bibnamefont{Tan}},
  \bibinfo{author}{\bibfnamefont{M.}~\bibnamefont{Wang}},
  \bibinfo{author}{\bibfnamefont{C.}~\bibnamefont{Zhang}},
  \bibinfo{author}{\bibfnamefont{P.}~\bibnamefont{Dai}}, \bibnamefont{and}
  \bibinfo{author}{\bibfnamefont{G.}~\bibnamefont{Blumberg}},
  \bibinfo{journal}{Phys. Rev. B} \textbf{\bibinfo{volume}{95}},
  \bibinfo{pages}{085125} (\bibinfo{year}{2017}).

\bibitem[{\citenamefont{B\"ohmer et~al.}(2014)\citenamefont{B\"ohmer, Burger,
  Hardy, Wolf, Schweiss, Fromknecht, Reinecker, Schranz, and
  Meingast}}]{bohmer2014}
\bibinfo{author}{\bibfnamefont{A.~E.} \bibnamefont{B\"ohmer}},
  \bibinfo{author}{\bibfnamefont{P.}~\bibnamefont{Burger}},
  \bibinfo{author}{\bibfnamefont{F.}~\bibnamefont{Hardy}},
  \bibinfo{author}{\bibfnamefont{T.}~\bibnamefont{Wolf}},
  \bibinfo{author}{\bibfnamefont{P.}~\bibnamefont{Schweiss}},
  \bibinfo{author}{\bibfnamefont{R.}~\bibnamefont{Fromknecht}},
  \bibinfo{author}{\bibfnamefont{M.}~\bibnamefont{Reinecker}},
  \bibinfo{author}{\bibfnamefont{W.}~\bibnamefont{Schranz}}, \bibnamefont{and}
  \bibinfo{author}{\bibfnamefont{C.}~\bibnamefont{Meingast}},
  \bibinfo{journal}{Phys. Rev. Lett.} \textbf{\bibinfo{volume}{112}},
  \bibinfo{pages}{047001} (\bibinfo{year}{2014}).

\bibitem[{\citenamefont{Frandsen et~al.}(2017)\citenamefont{Frandsen, Taddei,
  Yi, Frano, Guguchia, Yu, Si, Bugaris, Stadel, Osborn et~al.}}]{ben2017}
\bibinfo{author}{\bibfnamefont{B.~A.} \bibnamefont{Frandsen}},
  \bibinfo{author}{\bibfnamefont{K.~M.} \bibnamefont{Taddei}},
  \bibinfo{author}{\bibfnamefont{M.}~\bibnamefont{Yi}},
  \bibinfo{author}{\bibfnamefont{A.}~\bibnamefont{Frano}},
  \bibinfo{author}{\bibfnamefont{Z.}~\bibnamefont{Guguchia}},
  \bibinfo{author}{\bibfnamefont{R.}~\bibnamefont{Yu}},
  \bibinfo{author}{\bibfnamefont{Q.}~\bibnamefont{Si}},
  \bibinfo{author}{\bibfnamefont{D.~E.} \bibnamefont{Bugaris}},
  \bibinfo{author}{\bibfnamefont{R.}~\bibnamefont{Stadel}},
  \bibinfo{author}{\bibfnamefont{R.}~\bibnamefont{Osborn}},
  \bibnamefont{et~al.}, \bibinfo{journal}{Phys. Rev. Lett.}
  \textbf{\bibinfo{volume}{119}}, \bibinfo{pages}{187001}
  (\bibinfo{year}{2017}).

\bibitem[{\citenamefont{Wang et~al.}(2018{\natexlab{b}})\citenamefont{Wang, He,
  Hardy, Adelmann, Wolf, Merz, Schweiss, and Meingast}}]{LWang2018}
\bibinfo{author}{\bibfnamefont{L.}~\bibnamefont{Wang}},
  \bibinfo{author}{\bibfnamefont{M.}~\bibnamefont{He}},
  \bibinfo{author}{\bibfnamefont{F.}~\bibnamefont{Hardy}},
  \bibinfo{author}{\bibfnamefont{P.}~\bibnamefont{Adelmann}},
  \bibinfo{author}{\bibfnamefont{T.}~\bibnamefont{Wolf}},
  \bibinfo{author}{\bibfnamefont{M.}~\bibnamefont{Merz}},
  \bibinfo{author}{\bibfnamefont{P.}~\bibnamefont{Schweiss}}, \bibnamefont{and}
  \bibinfo{author}{\bibfnamefont{C.}~\bibnamefont{Meingast}},
  \bibinfo{journal}{Phys. Rev. B} \textbf{\bibinfo{volume}{97}},
  \bibinfo{pages}{224518} (\bibinfo{year}{2018}{\natexlab{b}}).

\bibitem[{\citenamefont{Niedziela et~al.}(2011)\citenamefont{Niedziela,
  Parshall, Lokshin, Sefat, Alatas, and Egami}}]{niedziela2011}
\bibinfo{author}{\bibfnamefont{J.~L.} \bibnamefont{Niedziela}},
  \bibinfo{author}{\bibfnamefont{D.}~\bibnamefont{Parshall}},
  \bibinfo{author}{\bibfnamefont{K.~A.} \bibnamefont{Lokshin}},
  \bibinfo{author}{\bibfnamefont{A.~S.} \bibnamefont{Sefat}},
  \bibinfo{author}{\bibfnamefont{A.}~\bibnamefont{Alatas}}, \bibnamefont{and}
  \bibinfo{author}{\bibfnamefont{T.}~\bibnamefont{Egami}},
  \bibinfo{journal}{Phys. Rev. B} \textbf{\bibinfo{volume}{84}},
  \bibinfo{pages}{224305} (\bibinfo{year}{2011}).

\bibitem[{\citenamefont{Parshall et~al.}(2015)\citenamefont{Parshall,
  Pintschovius, Niedziela, Castellan, Lamago, Mittal, Wolf, and
  Reznik}}]{parshall2015}
\bibinfo{author}{\bibfnamefont{D.}~\bibnamefont{Parshall}},
  \bibinfo{author}{\bibfnamefont{L.}~\bibnamefont{Pintschovius}},
  \bibinfo{author}{\bibfnamefont{J.~L.} \bibnamefont{Niedziela}},
  \bibinfo{author}{\bibfnamefont{J.~P.} \bibnamefont{Castellan}},
  \bibinfo{author}{\bibfnamefont{D.}~\bibnamefont{Lamago}},
  \bibinfo{author}{\bibfnamefont{R.}~\bibnamefont{Mittal}},
  \bibinfo{author}{\bibfnamefont{T.}~\bibnamefont{Wolf}}, \bibnamefont{and}
  \bibinfo{author}{\bibfnamefont{D.}~\bibnamefont{Reznik}},
  \bibinfo{journal}{Physical Review. B, Condensed Matter and Materials Physics}
  \textbf{\bibinfo{volume}{91}} (\bibinfo{year}{2015}), ISSN
  \bibinfo{issn}{1098-0121}.

\bibitem[{\citenamefont{Li et~al.}(2018)\citenamefont{Li, Yamani, Song, Wang,
  Zhang, Tam, Chen, Hu, Xu, Chi et~al.}}]{liyu2018}
\bibinfo{author}{\bibfnamefont{Y.}~\bibnamefont{Li}},
  \bibinfo{author}{\bibfnamefont{Z.}~\bibnamefont{Yamani}},
  \bibinfo{author}{\bibfnamefont{Y.}~\bibnamefont{Song}},
  \bibinfo{author}{\bibfnamefont{W.}~\bibnamefont{Wang}},
  \bibinfo{author}{\bibfnamefont{C.}~\bibnamefont{Zhang}},
  \bibinfo{author}{\bibfnamefont{D.~W.} \bibnamefont{Tam}},
  \bibinfo{author}{\bibfnamefont{T.}~\bibnamefont{Chen}},
  \bibinfo{author}{\bibfnamefont{D.}~\bibnamefont{Hu}},
  \bibinfo{author}{\bibfnamefont{Z.}~\bibnamefont{Xu}},
  \bibinfo{author}{\bibfnamefont{S.}~\bibnamefont{Chi}}, \bibnamefont{et~al.},
  \bibinfo{journal}{Phys. Rev. X} \textbf{\bibinfo{volume}{8}},
  \bibinfo{pages}{021056} (\bibinfo{year}{2018}).

\bibitem[{\citenamefont{Weber et~al.}(2018)\citenamefont{Weber, Parshall,
  Pintschovius, Castellan, Kauth, Merz, Wolf, Sch\"utt, Schmalian, Fernandes
  et~al.}}]{weber2018}
\bibinfo{author}{\bibfnamefont{F.}~\bibnamefont{Weber}},
  \bibinfo{author}{\bibfnamefont{D.}~\bibnamefont{Parshall}},
  \bibinfo{author}{\bibfnamefont{L.}~\bibnamefont{Pintschovius}},
  \bibinfo{author}{\bibfnamefont{J.-P.} \bibnamefont{Castellan}},
  \bibinfo{author}{\bibfnamefont{M.}~\bibnamefont{Kauth}},
  \bibinfo{author}{\bibfnamefont{M.}~\bibnamefont{Merz}},
  \bibinfo{author}{\bibfnamefont{T.}~\bibnamefont{Wolf}},
  \bibinfo{author}{\bibfnamefont{M.}~\bibnamefont{Sch\"utt}},
  \bibinfo{author}{\bibfnamefont{J.}~\bibnamefont{Schmalian}},
  \bibinfo{author}{\bibfnamefont{R.~M.} \bibnamefont{Fernandes}},
  \bibnamefont{et~al.}, \bibinfo{journal}{Phys. Rev. B}
  \textbf{\bibinfo{volume}{98}}, \bibinfo{pages}{014516}
  (\bibinfo{year}{2018}).

\bibitem[{\citenamefont{Merritt et~al.}(2020)\citenamefont{Merritt, Weber,
  Castellan, Wolf, Ishikawa, Said, Alatas, Fernandes, Baron, and
  Reznik}}]{merritt2020}
\bibinfo{author}{\bibfnamefont{A.~M.} \bibnamefont{Merritt}},
  \bibinfo{author}{\bibfnamefont{F.}~\bibnamefont{Weber}},
  \bibinfo{author}{\bibfnamefont{J.-P.} \bibnamefont{Castellan}},
  \bibinfo{author}{\bibfnamefont{T.}~\bibnamefont{Wolf}},
  \bibinfo{author}{\bibfnamefont{D.}~\bibnamefont{Ishikawa}},
  \bibinfo{author}{\bibfnamefont{A.~H.} \bibnamefont{Said}},
  \bibinfo{author}{\bibfnamefont{A.}~\bibnamefont{Alatas}},
  \bibinfo{author}{\bibfnamefont{R.~M.} \bibnamefont{Fernandes}},
  \bibinfo{author}{\bibfnamefont{A.~Q.~R.} \bibnamefont{Baron}},
  \bibnamefont{and} \bibinfo{author}{\bibfnamefont{D.}~\bibnamefont{Reznik}},
  \bibinfo{journal}{Phys. Rev. Lett.} \textbf{\bibinfo{volume}{124}},
  \bibinfo{pages}{157001} (\bibinfo{year}{2020}).

\bibitem[{\citenamefont{Frandsen et~al.}(2018)\citenamefont{Frandsen, Taddei,
  Bugaris, Stadel, Yi, Acharya, Osborn, Rosenkranz, Chmaissem, and
  Birgeneau}}]{ben2018}
\bibinfo{author}{\bibfnamefont{B.~A.} \bibnamefont{Frandsen}},
  \bibinfo{author}{\bibfnamefont{K.~M.} \bibnamefont{Taddei}},
  \bibinfo{author}{\bibfnamefont{D.~E.} \bibnamefont{Bugaris}},
  \bibinfo{author}{\bibfnamefont{R.}~\bibnamefont{Stadel}},
  \bibinfo{author}{\bibfnamefont{M.}~\bibnamefont{Yi}},
  \bibinfo{author}{\bibfnamefont{A.}~\bibnamefont{Acharya}},
  \bibinfo{author}{\bibfnamefont{R.}~\bibnamefont{Osborn}},
  \bibinfo{author}{\bibfnamefont{S.}~\bibnamefont{Rosenkranz}},
  \bibinfo{author}{\bibfnamefont{O.}~\bibnamefont{Chmaissem}},
  \bibnamefont{and} \bibinfo{author}{\bibfnamefont{R.~J.}
  \bibnamefont{Birgeneau}}, \bibinfo{journal}{Phys. Rev. B}
  \textbf{\bibinfo{volume}{98}}, \bibinfo{pages}{180505}
  (\bibinfo{year}{2018}).

\bibitem[{\citenamefont{Cowley}(1976)}]{cowley1976}
\bibinfo{author}{\bibfnamefont{R.~A.} \bibnamefont{Cowley}},
  \bibinfo{journal}{Phys. Rev. B} \textbf{\bibinfo{volume}{13}},
  \bibinfo{pages}{4877} (\bibinfo{year}{1976}).

\bibitem[{\citenamefont{{Als-Nielsen} and {Birgeneau}}(1977)}]{nielsen1977}
\bibinfo{author}{\bibfnamefont{J.}~\bibnamefont{{Als-Nielsen}}}
  \bibnamefont{and} \bibinfo{author}{\bibfnamefont{R.~J.}
  \bibnamefont{{Birgeneau}}}, \bibinfo{journal}{American Journal of Physics}
  \textbf{\bibinfo{volume}{45}}, \bibinfo{pages}{554} (\bibinfo{year}{1977}).

\bibitem[{\citenamefont{Karahasanovic and Schmalian}(2016)}]{karahsanovic2016}
\bibinfo{author}{\bibfnamefont{U.}~\bibnamefont{Karahasanovic}}
  \bibnamefont{and}
  \bibinfo{author}{\bibfnamefont{J.}~\bibnamefont{Schmalian}},
  \bibinfo{journal}{Phys. Rev. B} \textbf{\bibinfo{volume}{93}},
  \bibinfo{pages}{064520} (\bibinfo{year}{2016}).

\bibitem[{\citenamefont{Baron et~al.}(2000)\citenamefont{Baron, Tanaka, Goto,
  Takeshita, Matsushita, and Ishikawa}}]{baron}
\bibinfo{author}{\bibfnamefont{A.}~\bibnamefont{Baron}},
  \bibinfo{author}{\bibfnamefont{Y.}~\bibnamefont{Tanaka}},
  \bibinfo{author}{\bibfnamefont{S.}~\bibnamefont{Goto}},
  \bibinfo{author}{\bibfnamefont{K.}~\bibnamefont{Takeshita}},
  \bibinfo{author}{\bibfnamefont{T.}~\bibnamefont{Matsushita}},
  \bibnamefont{and} \bibinfo{author}{\bibfnamefont{T.}~\bibnamefont{Ishikawa}},
  \bibinfo{journal}{Journal of Physics and Chemistry of Solids}
  \textbf{\bibinfo{volume}{61}}, \bibinfo{pages}{461 } (\bibinfo{year}{2000}),
  ISSN \bibinfo{issn}{0022-3697}.

\bibitem[{\citenamefont{Toellner et~al.}(2011)\citenamefont{Toellner, Alatas,
  and Said}}]{Toellner}
\bibinfo{author}{\bibfnamefont{T.~S.} \bibnamefont{Toellner}},
  \bibinfo{author}{\bibfnamefont{A.}~\bibnamefont{Alatas}}, \bibnamefont{and}
  \bibinfo{author}{\bibfnamefont{A.~H.} \bibnamefont{Said}},
  \bibinfo{journal}{Journal of Synchrotron Radiation}
  \textbf{\bibinfo{volume}{18}}, \bibinfo{pages}{605} (\bibinfo{year}{2011}).

\bibitem[{\citenamefont{Said et~al.}(2020)\citenamefont{Said, Sinn, Toellner,
  Alp, Gog, Leu, Bean, and Alatas}}]{Said}
\bibinfo{author}{\bibfnamefont{A.~H.} \bibnamefont{Said}},
  \bibinfo{author}{\bibfnamefont{H.}~\bibnamefont{Sinn}},
  \bibinfo{author}{\bibfnamefont{T.~S.} \bibnamefont{Toellner}},
  \bibinfo{author}{\bibfnamefont{E.~E.} \bibnamefont{Alp}},
  \bibinfo{author}{\bibfnamefont{T.}~\bibnamefont{Gog}},
  \bibinfo{author}{\bibfnamefont{B.~M.} \bibnamefont{Leu}},
  \bibinfo{author}{\bibfnamefont{S.}~\bibnamefont{Bean}}, \bibnamefont{and}
  \bibinfo{author}{\bibfnamefont{A.}~\bibnamefont{Alatas}},
  \bibinfo{journal}{Journal of Synchrotron Radiation}
  \textbf{\bibinfo{volume}{27}}, \bibinfo{pages}{827} (\bibinfo{year}{2020}).

\bibitem[{SI()}]{SI}
\bibinfo{note}{See Supplemental Material for details on experimental setup,
  extraction of phonon energies, extraction of the nematic correlation length,
  comparison of $\xi_0$ between different systems and a search for phonon
  back-folding in the AFM-T phase.}

\bibitem[{\citenamefont{Yu et~al.}(2017)\citenamefont{Yu, Yi, Frandsen,
  Birgeneau, and Si}}]{RYu2017}
\bibinfo{author}{\bibfnamefont{R.}~\bibnamefont{Yu}},
  \bibinfo{author}{\bibfnamefont{M.}~\bibnamefont{Yi}},
  \bibinfo{author}{\bibfnamefont{B.~A.} \bibnamefont{Frandsen}},
  \bibinfo{author}{\bibfnamefont{R.~J.} \bibnamefont{Birgeneau}},
  \bibnamefont{and} \bibinfo{author}{\bibfnamefont{Q.}~\bibnamefont{Si}},
  \emph{\bibinfo{title}{Emergent phases in iron pnictides: Double-q
  antiferromagnetism, charge order and enhanced nematic correlations}}
  (\bibinfo{year}{2017}), \eprint{arXiv:1706.07087}.

\bibitem[{\citenamefont{Dai}(2015)}]{PDai2015}
\bibinfo{author}{\bibfnamefont{P.}~\bibnamefont{Dai}}, \bibinfo{journal}{Rev.
  Mod. Phys.} \textbf{\bibinfo{volume}{87}}, \bibinfo{pages}{855}
  (\bibinfo{year}{2015}).

\bibitem[{\citenamefont{Grinenko et~al.}(2020)\citenamefont{Grinenko, Sarkar,
  Kihou, Lee, Morozov, Aswartham, B\"{u}chner, Chekhonin, Skrotzki, Nenkov
  et~al.}}]{Grinenko2020}
\bibinfo{author}{\bibfnamefont{V.}~\bibnamefont{Grinenko}},
  \bibinfo{author}{\bibfnamefont{R.}~\bibnamefont{Sarkar}},
  \bibinfo{author}{\bibfnamefont{K.}~\bibnamefont{Kihou}},
  \bibinfo{author}{\bibfnamefont{C.~H.} \bibnamefont{Lee}},
  \bibinfo{author}{\bibfnamefont{I.}~\bibnamefont{Morozov}},
  \bibinfo{author}{\bibfnamefont{S.}~\bibnamefont{Aswartham}},
  \bibinfo{author}{\bibfnamefont{B.}~\bibnamefont{B\"{u}chner}},
  \bibinfo{author}{\bibfnamefont{P.}~\bibnamefont{Chekhonin}},
  \bibinfo{author}{\bibfnamefont{W.}~\bibnamefont{Skrotzki}},
  \bibinfo{author}{\bibfnamefont{K.}~\bibnamefont{Nenkov}},
  \bibnamefont{et~al.}, \bibinfo{journal}{Nature Physics}
  \textbf{\bibinfo{volume}{16}}, \bibinfo{pages}{789} (\bibinfo{year}{2020}).

\bibitem[{\citenamefont{Song et~al.}(2016)\citenamefont{Song, Man, Zhang, Lu,
  Zhang, Wang, Tan, Regnault, Su, Kang et~al.}}]{YSong2016}
\bibinfo{author}{\bibfnamefont{Y.}~\bibnamefont{Song}},
  \bibinfo{author}{\bibfnamefont{H.}~\bibnamefont{Man}},
  \bibinfo{author}{\bibfnamefont{R.}~\bibnamefont{Zhang}},
  \bibinfo{author}{\bibfnamefont{X.}~\bibnamefont{Lu}},
  \bibinfo{author}{\bibfnamefont{C.}~\bibnamefont{Zhang}},
  \bibinfo{author}{\bibfnamefont{M.}~\bibnamefont{Wang}},
  \bibinfo{author}{\bibfnamefont{G.}~\bibnamefont{Tan}},
  \bibinfo{author}{\bibfnamefont{L.-P.} \bibnamefont{Regnault}},
  \bibinfo{author}{\bibfnamefont{Y.}~\bibnamefont{Su}},
  \bibinfo{author}{\bibfnamefont{J.}~\bibnamefont{Kang}}, \bibnamefont{et~al.},
  \bibinfo{journal}{Phys. Rev. B} \textbf{\bibinfo{volume}{94}},
  \bibinfo{pages}{214516} (\bibinfo{year}{2016}).

\bibitem[{\citenamefont{Guo et~al.}(2019)\citenamefont{Guo, Yue, Iida,
  Kamazawa, Chen, Han, Zhang, and Li}}]{yuan2019}
\bibinfo{author}{\bibfnamefont{J.}~\bibnamefont{Guo}},
  \bibinfo{author}{\bibfnamefont{L.}~\bibnamefont{Yue}},
  \bibinfo{author}{\bibfnamefont{K.}~\bibnamefont{Iida}},
  \bibinfo{author}{\bibfnamefont{K.}~\bibnamefont{Kamazawa}},
  \bibinfo{author}{\bibfnamefont{L.}~\bibnamefont{Chen}},
  \bibinfo{author}{\bibfnamefont{T.}~\bibnamefont{Han}},
  \bibinfo{author}{\bibfnamefont{Y.}~\bibnamefont{Zhang}}, \bibnamefont{and}
  \bibinfo{author}{\bibfnamefont{Y.}~\bibnamefont{Li}}, \bibinfo{journal}{Phys.
  Rev. Lett.} \textbf{\bibinfo{volume}{122}}, \bibinfo{pages}{017001}
  (\bibinfo{year}{2019}).

\bibitem[{\citenamefont{Christensen et~al.}(2015)\citenamefont{Christensen,
  Kang, Andersen, Eremin, and Fernandes}}]{christensen2015}
\bibinfo{author}{\bibfnamefont{M.~H.} \bibnamefont{Christensen}},
  \bibinfo{author}{\bibfnamefont{J.}~\bibnamefont{Kang}},
  \bibinfo{author}{\bibfnamefont{B.~M.} \bibnamefont{Andersen}},
  \bibinfo{author}{\bibfnamefont{I.}~\bibnamefont{Eremin}}, \bibnamefont{and}
  \bibinfo{author}{\bibfnamefont{R.~M.} \bibnamefont{Fernandes}},
  \bibinfo{journal}{Phys. Rev. B} \textbf{\bibinfo{volume}{92}},
  \bibinfo{pages}{214509} (\bibinfo{year}{2015}).

\bibitem[{\citenamefont{Timmons et~al.}(2019)\citenamefont{Timmons, Tanatar,
  Willa, Teknowijoyo, Cho, Ko\ifmmode~\acute{n}\else \'{n}\fi{}czykowski,
  Cavani, Liu, Lograsso, Welp et~al.}}]{timmons2019}
\bibinfo{author}{\bibfnamefont{E.~I.} \bibnamefont{Timmons}},
  \bibinfo{author}{\bibfnamefont{M.~A.} \bibnamefont{Tanatar}},
  \bibinfo{author}{\bibfnamefont{K.}~\bibnamefont{Willa}},
  \bibinfo{author}{\bibfnamefont{S.}~\bibnamefont{Teknowijoyo}},
  \bibinfo{author}{\bibfnamefont{K.}~\bibnamefont{Cho}},
  \bibinfo{author}{\bibfnamefont{M.}~\bibnamefont{Ko\ifmmode~\acute{n}\else
  \'{n}\fi{}czykowski}},
  \bibinfo{author}{\bibfnamefont{O.}~\bibnamefont{Cavani}},
  \bibinfo{author}{\bibfnamefont{Y.}~\bibnamefont{Liu}},
  \bibinfo{author}{\bibfnamefont{T.~A.} \bibnamefont{Lograsso}},
  \bibinfo{author}{\bibfnamefont{U.}~\bibnamefont{Welp}}, \bibnamefont{et~al.},
  \bibinfo{journal}{Phys. Rev. B} \textbf{\bibinfo{volume}{99}},
  \bibinfo{pages}{054518} (\bibinfo{year}{2019}).

\bibitem[{\citenamefont{Hoyer et~al.}(2016)\citenamefont{Hoyer, Fernandes,
  Levchenko, and Schmalian}}]{hoyer2016}
\bibinfo{author}{\bibfnamefont{M.}~\bibnamefont{Hoyer}},
  \bibinfo{author}{\bibfnamefont{R.~M.} \bibnamefont{Fernandes}},
  \bibinfo{author}{\bibfnamefont{A.}~\bibnamefont{Levchenko}},
  \bibnamefont{and}
  \bibinfo{author}{\bibfnamefont{J.}~\bibnamefont{Schmalian}},
  \bibinfo{journal}{Phys. Rev. B} \textbf{\bibinfo{volume}{93}},
  \bibinfo{pages}{144414} (\bibinfo{year}{2016}).

\bibitem[{\citenamefont{Gastiasoro and Andersen}(2015)}]{gastiasoro2015}
\bibinfo{author}{\bibfnamefont{M.~N.} \bibnamefont{Gastiasoro}}
  \bibnamefont{and} \bibinfo{author}{\bibfnamefont{B.~M.}
  \bibnamefont{Andersen}}, \bibinfo{journal}{Phys. Rev. B}
  \textbf{\bibinfo{volume}{92}}, \bibinfo{pages}{140506}
  (\bibinfo{year}{2015}).

\bibitem[{\citenamefont{Hardy et~al.}(2016)\citenamefont{Hardy, B\"ohmer, de'
  Medici, Capone, Giovannetti, Eder, Wang, He, Wolf, Schweiss
  et~al.}}]{hardy2016}
\bibinfo{author}{\bibfnamefont{F.}~\bibnamefont{Hardy}},
  \bibinfo{author}{\bibfnamefont{A.~E.} \bibnamefont{B\"ohmer}},
  \bibinfo{author}{\bibfnamefont{L.}~\bibnamefont{de' Medici}},
  \bibinfo{author}{\bibfnamefont{M.}~\bibnamefont{Capone}},
  \bibinfo{author}{\bibfnamefont{G.}~\bibnamefont{Giovannetti}},
  \bibinfo{author}{\bibfnamefont{R.}~\bibnamefont{Eder}},
  \bibinfo{author}{\bibfnamefont{L.}~\bibnamefont{Wang}},
  \bibinfo{author}{\bibfnamefont{M.}~\bibnamefont{He}},
  \bibinfo{author}{\bibfnamefont{T.}~\bibnamefont{Wolf}},
  \bibinfo{author}{\bibfnamefont{P.}~\bibnamefont{Schweiss}},
  \bibnamefont{et~al.}, \bibinfo{journal}{Phys. Rev. B}
  \textbf{\bibinfo{volume}{94}}, \bibinfo{pages}{205113}
  (\bibinfo{year}{2016}).

\bibitem[{\citenamefont{Pratt et~al.}(2009)\citenamefont{Pratt, Tian, Kreyssig,
  Zarestky, Nandi, Ni, Bud'ko, Canfield, Goldman, and McQueeney}}]{DPratt2009}
\bibinfo{author}{\bibfnamefont{D.~K.} \bibnamefont{Pratt}},
  \bibinfo{author}{\bibfnamefont{W.}~\bibnamefont{Tian}},
  \bibinfo{author}{\bibfnamefont{A.}~\bibnamefont{Kreyssig}},
  \bibinfo{author}{\bibfnamefont{J.~L.} \bibnamefont{Zarestky}},
  \bibinfo{author}{\bibfnamefont{S.}~\bibnamefont{Nandi}},
  \bibinfo{author}{\bibfnamefont{N.}~\bibnamefont{Ni}},
  \bibinfo{author}{\bibfnamefont{S.~L.} \bibnamefont{Bud'ko}},
  \bibinfo{author}{\bibfnamefont{P.~C.} \bibnamefont{Canfield}},
  \bibinfo{author}{\bibfnamefont{A.~I.} \bibnamefont{Goldman}},
  \bibnamefont{and} \bibinfo{author}{\bibfnamefont{R.~J.}
  \bibnamefont{McQueeney}}, \bibinfo{journal}{Phys. Rev. Lett.}
  \textbf{\bibinfo{volume}{103}}, \bibinfo{pages}{087001}
  (\bibinfo{year}{2009}).

\bibitem[{\citenamefont{Nandi et~al.}(2010)\citenamefont{Nandi, Kim, Kreyssig,
  Fernandes, Pratt, Thaler, Ni, Bud'ko, Canfield, Schmalian
  et~al.}}]{SNandi2010}
\bibinfo{author}{\bibfnamefont{S.}~\bibnamefont{Nandi}},
  \bibinfo{author}{\bibfnamefont{M.~G.} \bibnamefont{Kim}},
  \bibinfo{author}{\bibfnamefont{A.}~\bibnamefont{Kreyssig}},
  \bibinfo{author}{\bibfnamefont{R.~M.} \bibnamefont{Fernandes}},
  \bibinfo{author}{\bibfnamefont{D.~K.} \bibnamefont{Pratt}},
  \bibinfo{author}{\bibfnamefont{A.}~\bibnamefont{Thaler}},
  \bibinfo{author}{\bibfnamefont{N.}~\bibnamefont{Ni}},
  \bibinfo{author}{\bibfnamefont{S.~L.} \bibnamefont{Bud'ko}},
  \bibinfo{author}{\bibfnamefont{P.~C.} \bibnamefont{Canfield}},
  \bibinfo{author}{\bibfnamefont{J.}~\bibnamefont{Schmalian}},
  \bibnamefont{et~al.}, \bibinfo{journal}{Phys. Rev. Lett.}
  \textbf{\bibinfo{volume}{104}}, \bibinfo{pages}{057006}
  (\bibinfo{year}{2010}).

\bibitem[{\citenamefont{Yi et~al.}(2014)\citenamefont{Yi, Zhang, Liu, Ding,
  Chu, Kemper, Plonka, Moritz, Hashimoto, Mo et~al.}}]{Yi2014}
\bibinfo{author}{\bibfnamefont{M.}~\bibnamefont{Yi}},
  \bibinfo{author}{\bibfnamefont{Y.}~\bibnamefont{Zhang}},
  \bibinfo{author}{\bibfnamefont{Z.-K.} \bibnamefont{Liu}},
  \bibinfo{author}{\bibfnamefont{X.}~\bibnamefont{Ding}},
  \bibinfo{author}{\bibfnamefont{J.-H.} \bibnamefont{Chu}},
  \bibinfo{author}{\bibfnamefont{A.}~\bibnamefont{Kemper}},
  \bibinfo{author}{\bibfnamefont{N.}~\bibnamefont{Plonka}},
  \bibinfo{author}{\bibfnamefont{B.}~\bibnamefont{Moritz}},
  \bibinfo{author}{\bibfnamefont{M.}~\bibnamefont{Hashimoto}},
  \bibinfo{author}{\bibfnamefont{S.-K.} \bibnamefont{Mo}},
  \bibnamefont{et~al.}, \bibinfo{journal}{Nature Communications}
  \textbf{\bibinfo{volume}{5}} (\bibinfo{year}{2014}).

\bibitem[{\citenamefont{Lederer et~al.}(2015)\citenamefont{Lederer, Schattner,
  Berg, and Kivelson}}]{PhysRevLett.114.097001}
\bibinfo{author}{\bibfnamefont{S.}~\bibnamefont{Lederer}},
  \bibinfo{author}{\bibfnamefont{Y.}~\bibnamefont{Schattner}},
  \bibinfo{author}{\bibfnamefont{E.}~\bibnamefont{Berg}}, \bibnamefont{and}
  \bibinfo{author}{\bibfnamefont{S.~A.} \bibnamefont{Kivelson}},
  \bibinfo{journal}{Phys. Rev. Lett.} \textbf{\bibinfo{volume}{114}},
  \bibinfo{pages}{097001} (\bibinfo{year}{2015}),
  \urlprefix\url{https://link.aps.org/doi/10.1103/PhysRevLett.114.097001}.

\bibitem[{\citenamefont{Fernandes and Millis}(2013)}]{fernandes2013}
\bibinfo{author}{\bibfnamefont{R.~M.} \bibnamefont{Fernandes}}
  \bibnamefont{and} \bibinfo{author}{\bibfnamefont{A.~J.}
  \bibnamefont{Millis}}, \bibinfo{journal}{Phys. Rev. Lett.}
  \textbf{\bibinfo{volume}{111}}, \bibinfo{pages}{127001}
  (\bibinfo{year}{2013}).

\bibitem[{\citenamefont{Wilson et~al.}(2010)\citenamefont{Wilson, Yamani,
  Rotundu, Freelon, Valdivia, Bourret-Courchesne, Lynn, Chi, Hong, and
  Birgeneau}}]{SWilson2010}
\bibinfo{author}{\bibfnamefont{S.~D.} \bibnamefont{Wilson}},
  \bibinfo{author}{\bibfnamefont{Z.}~\bibnamefont{Yamani}},
  \bibinfo{author}{\bibfnamefont{C.~R.} \bibnamefont{Rotundu}},
  \bibinfo{author}{\bibfnamefont{B.}~\bibnamefont{Freelon}},
  \bibinfo{author}{\bibfnamefont{P.~N.} \bibnamefont{Valdivia}},
  \bibinfo{author}{\bibfnamefont{E.}~\bibnamefont{Bourret-Courchesne}},
  \bibinfo{author}{\bibfnamefont{J.~W.} \bibnamefont{Lynn}},
  \bibinfo{author}{\bibfnamefont{S.}~\bibnamefont{Chi}},
  \bibinfo{author}{\bibfnamefont{T.}~\bibnamefont{Hong}}, \bibnamefont{and}
  \bibinfo{author}{\bibfnamefont{R.~J.} \bibnamefont{Birgeneau}},
  \bibinfo{journal}{Phys. Rev. B} \textbf{\bibinfo{volume}{82}},
  \bibinfo{pages}{144502} (\bibinfo{year}{2010}).

\bibitem[{\citenamefont{Altarawneh et~al.}(2008)\citenamefont{Altarawneh,
  Collar, Mielke, Ni, Bud'ko, and Canfield}}]{MAltarawneh2008}
\bibinfo{author}{\bibfnamefont{M.~M.} \bibnamefont{Altarawneh}},
  \bibinfo{author}{\bibfnamefont{K.}~\bibnamefont{Collar}},
  \bibinfo{author}{\bibfnamefont{C.~H.} \bibnamefont{Mielke}},
  \bibinfo{author}{\bibfnamefont{N.}~\bibnamefont{Ni}},
  \bibinfo{author}{\bibfnamefont{S.~L.} \bibnamefont{Bud'ko}},
  \bibnamefont{and} \bibinfo{author}{\bibfnamefont{P.~C.}
  \bibnamefont{Canfield}}, \bibinfo{journal}{Phys. Rev. B}
  \textbf{\bibinfo{volume}{78}}, \bibinfo{pages}{220505}
  (\bibinfo{year}{2008}).

\bibitem[{\citenamefont{Yuan et~al.}(2009)\citenamefont{Yuan, Singleton,
  Balakirev, Baily, Chen, Luo, and Wang}}]{Yuan2009}
\bibinfo{author}{\bibfnamefont{H.~Q.} \bibnamefont{Yuan}},
  \bibinfo{author}{\bibfnamefont{J.}~\bibnamefont{Singleton}},
  \bibinfo{author}{\bibfnamefont{F.~F.} \bibnamefont{Balakirev}},
  \bibinfo{author}{\bibfnamefont{S.~A.} \bibnamefont{Baily}},
  \bibinfo{author}{\bibfnamefont{G.~F.} \bibnamefont{Chen}},
  \bibinfo{author}{\bibfnamefont{J.~L.} \bibnamefont{Luo}}, \bibnamefont{and}
  \bibinfo{author}{\bibfnamefont{N.~L.} \bibnamefont{Wang}},
  \bibinfo{journal}{Nature} \textbf{\bibinfo{volume}{457}},
  \bibinfo{pages}{565} (\bibinfo{year}{2009}).

\bibitem[{\citenamefont{Loh et~al.}(2010)\citenamefont{Loh, Carlson, and
  Dahmen}}]{YLoh2010}
\bibinfo{author}{\bibfnamefont{Y.~L.} \bibnamefont{Loh}},
  \bibinfo{author}{\bibfnamefont{E.~W.} \bibnamefont{Carlson}},
  \bibnamefont{and} \bibinfo{author}{\bibfnamefont{K.~A.}
  \bibnamefont{Dahmen}}, \bibinfo{journal}{Phys. Rev. B}
  \textbf{\bibinfo{volume}{81}}, \bibinfo{pages}{224207}
  (\bibinfo{year}{2010}).

\bibitem[{\citenamefont{Carlson and Dahmen}(2011)}]{Carlson2011}
\bibinfo{author}{\bibfnamefont{E.}~\bibnamefont{Carlson}} \bibnamefont{and}
  \bibinfo{author}{\bibfnamefont{K.}~\bibnamefont{Dahmen}},
  \bibinfo{journal}{Nature Communications} \textbf{\bibinfo{volume}{2}}
  (\bibinfo{year}{2011}).

\bibitem[{\citenamefont{Zhang et~al.}(2019)\citenamefont{Zhang, Wei, Xie, Liu,
  Gong, Ma, Hu, \ifmmode~\check{C}\else \v{C}\fi{}erm\'ak, Schneidewind, Tucker
  et~al.}}]{zhang2019}
\bibinfo{author}{\bibfnamefont{W.}~\bibnamefont{Zhang}},
  \bibinfo{author}{\bibfnamefont{Y.}~\bibnamefont{Wei}},
  \bibinfo{author}{\bibfnamefont{T.}~\bibnamefont{Xie}},
  \bibinfo{author}{\bibfnamefont{Z.}~\bibnamefont{Liu}},
  \bibinfo{author}{\bibfnamefont{D.}~\bibnamefont{Gong}},
  \bibinfo{author}{\bibfnamefont{X.}~\bibnamefont{Ma}},
  \bibinfo{author}{\bibfnamefont{D.}~\bibnamefont{Hu}},
  \bibinfo{author}{\bibfnamefont{P.}~\bibnamefont{\ifmmode~\check{C}\else
  \v{C}\fi{}erm\'ak}},
  \bibinfo{author}{\bibfnamefont{A.}~\bibnamefont{Schneidewind}},
  \bibinfo{author}{\bibfnamefont{G.}~\bibnamefont{Tucker}},
  \bibnamefont{et~al.}, \bibinfo{journal}{Phys. Rev. Lett.}
  \textbf{\bibinfo{volume}{122}}, \bibinfo{pages}{037001}
  (\bibinfo{year}{2019}).

\end{thebibliography}
\end{document}